\documentclass[journal]{IEEEtran}

\usepackage[pdftex]{graphicx}
\graphicspath{{../pdf/}{../jpeg/}{../png/}}
\DeclareGraphicsExtensions{.pdf,.jpeg,.png}

\usepackage{amsmath}
\usepackage{mathtools}
\usepackage{array}
\usepackage{standalone}
\usepackage[normalem]{ulem}
\usepackage{etoolbox}
\newtoggle{arXiv}
\newtoggle{bibrun}
\usepackage{newtxtext, newtxmath}
\usepackage{graphicx}
\usepackage{color, calc}
\usepackage{centernot}
\usepackage{array, booktabs}
\newcolumntype{L}{>{$}l<{$}}
\newcolumntype{C}{>{$}c<{$}}
\newcolumntype{R}{>{$}r<{$}}
\usepackage{tikz, tikz-3dplot}
\usetikzlibrary{arrows, arrows.meta, calc, positioning}
\usetikzlibrary{decorations.pathmorphing}	
\tikzset{>=Stealth}
\usepackage{pgfplots}
\usepackage{siunitx, physics}
\newcommand{\I}{\ensuremath{\mathrm{i}}}

\usepackage{enumitem}
\setlist[description]{labelindent=0pt, leftmargin=\parindent, font=\normalfont\itshape}
\usepackage{url}
\usepackage{subfigure}
\usepackage{textcomp}
\usepackage{eurosym}
\usepackage{stfloats}
\usepackage{lipsum}
\usepackage{array}
\usepackage{colortbl}
\usepackage{multirow}
\usepackage{booktabs}
\usepackage{makecell}
\usepackage{hyperref}

\begin{document}
\renewcommand\arraystretch{1.5}

\title{Qualitative and quantitative hard-tissue MRI with portable Halbach scanners}

\author{\IEEEauthorblockN{
		Jose~Borreguero\IEEEauthorrefmark{1},
		Luiz~G.~C.~Santos\IEEEauthorrefmark{1},
		Lorena~Vega~Cid\IEEEauthorrefmark{1},
		Eli~G.~Castanon\IEEEauthorrefmark{1},
		Marina~Fernández-García\IEEEauthorrefmark{1},
		Pablo~Benlloch\IEEEauthorrefmark{1}$^{,}$\IEEEauthorrefmark{2},
		Rubén~Bosch\IEEEauthorrefmark{1},
		Jesús~Conejero\IEEEauthorrefmark{1},
		Pablo~García-Cristóbal\IEEEauthorrefmark{1},
		Alba~González-Cebrián\IEEEauthorrefmark{1},
		Teresa~Guallart-Naval\IEEEauthorrefmark{1},
		Eduardo~Pallás\IEEEauthorrefmark{1},
		Laia~Porcar\IEEEauthorrefmark{3},
		Lucas~Swistunow\IEEEauthorrefmark{1},
		Jose Miguel Algarín\IEEEauthorrefmark{1}$^{,}$\IEEEauthorrefmark{2},
		Fernando~Galve\IEEEauthorrefmark{1} and
		Joseba~Alonso\IEEEauthorrefmark{1}}
	
	\IEEEauthorblockA{\IEEEauthorrefmark{1}MRILab, Institute for Molecular Imaging and Instrumentation (i3M), Spanish National Research Council (CSIC) and Universitat Polit\`ecnica de Val\`encia (UPV), 46022 Valencia, Spain\\
	\IEEEauthorrefmark{2}Full Body Insight S.L., 46980 Paterna, Spain\\
	\IEEEauthorrefmark{3}PhysioMRI Tech S.L., 46980 Paterna, Spain}

\thanks{Corresponding author: J. Borreguero (pepe.morata@i3m.upv.es).}}

\maketitle

\IEEEtitleabstractindextext{%
\begin{abstract}
	\newline
	Purpose: {\normalfont To demonstrate the feasibility of performing \textit{in-vivo} imaging and quantitative relaxation mapping of soft and hard tissues using a low-cost, portable MRI scanner, and to establish the methodological foundations for zero echo time (ZTE) imaging in systems affected by strong field inhomogeneities.}\\
	Methods: {\normalfont A complete framework for artifact-free ZTE imaging at low field was developed, including: (i)~RF pulse pre/counter-emphasis calibration to minimize ring-down and electronics switching time; (ii)~an extension of a recent single-point double-shot (SPDS) protocol for simultaneous $B_0$ and $B_1$ mapping; and (iii)~a model-based reconstruction incorporating these field maps into the encoding matrix. ZTE imaging and variable flip angle (VFA) $T_1$ mapping were performed on phantoms and \textit{in-vivo} human knees and ankles, and benchmarked against standard RARE and STIR acquisitions.}\\
	Results: {\normalfont The optimized PETRA sequence produced 3D images of knees and ankles within clinically compatible times ($<15$\,min), revealing hard tissues such as ligaments, tendons, cartilage, and bone that are invisible in spin-echo sequences. The extended SPDS method enabled accurate field mapping, while the VFA approach provided the first \textit{in-vivo} $T_1$ measurements of hard tissues at $B_0<0.1$\,T.}\\
	Conclusions: {\normalfont The proposed framework broadens the range of pulse sequences feasible in portable low-field MRI and demonstrates the potential of ZTE for quantitative and structural imaging of musculoskeletal tissues in affordable Halbach-based systems.}
\end{abstract}

}

\maketitle

\IEEEdisplaynontitleabstractindextext

\section{Introduction}

In recent years, there has been growing interest in exploring the potential of low-field MRI (LF-MRI, $B_0<0.5$\,T) as a way to enhance accessibility and democratize medical imaging \cite{Sarracanie2015,Geethanath2019,Obungoloch2023,Wald2020,GuallartPortable2022}. The signal-to-noise ratio (SNR) deficit inherent to this regime is inevitable compared to clinical standards ($B_0=1.5$ and 3\,T), resulting in anatomical images of lower resolution than those obtained with conventional MRI. Nevertheless, a variety of strategies have been developed to make LF-MRI reconstructions diagnostically useful, including both hardware optimization \cite{WebbandReilly2023,OREILLY2019134} and software-based enhancement \cite{Ayde2025}. 

Particularly interesting are scanners employing a Halbach-array main magnet \cite{halbach1980design}, which can provide sufficient $B_0$ strength and homogeneity while substantially reducing the weight, footprint, and cost of MRI systems \cite{OReilly2020,Cooley2020,galve2024elliptical}. Halbach-based systems have shown potential in diverse scenarios, such as mounted on vehicles \cite{Nakagomi2019,Miyasaka2022}, at crowded sporting events \cite{algarin2024portable}, or even for home-based imaging \cite{GuallartPortable2022}.

Even so, the limited hardware capabilities of Halbach systems constrain the range of pulse sequences (and thereby clinical applications) that can be implemented \cite{OREILLY2019134}. The effects of the relatively poor $B_0$ homogeneity is alleviated through spin-echo refocusing sequences \cite{Hahn1950Echoes,Carr1954}, which mitigate the SNR loss. This encoding strategy is, however, only suitable for tissues with long transverse relaxation times ($T_2>20$\,ms), a condition typically met only by water-rich soft tissues.

However, a variety of tissues in the human body are characterized by short $T_2$ times (even $<1$\,ms), including bone, lungs, ligaments, tendons, and teeth \cite{Mastrogiacomo2019}. The NMR signal from these tissues is heavily attenuated and X-ray-based imaging techniques are favored. Nonetheless, the ability to directly visualize them using low-cost MRI could offer substantial diagnostic and therapeutic advantages over conventional radiography. Unlike X-rays, MRI enables innocuous and direct assessment of fiber continuity, thickness, and internal signal of tendons and ligaments, as well as early detection of subcortical bone marrow edema, microfractures, and cartilage degeneration \cite{Eck2022}.

To image these tissues, dedicated pulse sequences have been developed. These include ultra-short echo time (UTE) \cite{robson2003ultrashort}, zero echo time (ZTE) \cite{Madio1995}, or Sweep Imaging with Fourier Transformation (SWIFT) \cite{IDIYATULLIN2006342}. Their performance has been validated across scanners of varying $B_0$ and for diverse applications, such as imaging and diagnosis of pulmonary \cite{LIN2024e34098}, cartilaginous \cite{Afsahi}, ligamentous \cite{Malhi}, and dental lesions \cite{Weiger2012,Idiyatullin2011}, even at fields as low as 260\,mT \cite{Algarin2020}. In exchange, these sequences are generally hardware-demanding, requiring high $B_0$ homogeneity, substantial radio-frequency (RF) power, and strong gradients. Such requirements conflict with the capabilities of low-cost, Halbach-based systems, hence hard-tissue imaging remains unproven with portable scanners.

In this work, we present a proof-of-concept demonstration of the feasibility of applying ZTE sequences in a portable Halbach system originally designed for \emph{in-vivo} brain and musculoskeletal imaging with spin-echo acquisitions. To overcome signal losses associated with RF switching delays at low field, we implement a Pointwise Encoding Time Reduction with Radial Acquisition (PETRA) sequence \cite{Grodzki2012}. To compensate the limited hardware capabilities of portable Halbach systems, we introduce model-based reconstructions that incorporate prior knowledge (PK) of the main ($B_0$) and RF ($B_1$) field inhomogeneities. This PK is obtained via an extension of the Single-Point Double-Shot (SPDS) protocol \cite{Borreguero2024}. In addition, we extend PETRA to generate \emph{in-vivo} quantitative spin-lattice relaxation ($T_1$) maps using a variable flip-angle approach (VFA-PETRA) under incoherent steady-state conditions. This strategy enables the simultaneous quantification of $T_1$ times in muscle and fat, as well as structures typically invisible to MRI, including tendons, ligaments, and cortical bone.

\section{Background \& theory}
\label{sec:theory}

\subsection{Relaxation and dead times}
\label{sec:times}

Some aspects should be considered when advancing towards ZTE with Halbach systems. One of the key challenges is the available time margin for NMR signal acquisition. In ZTE, the receive (Rx) window should start only after switching the RF electronics from transmit (Tx) to Rx mode. This dead time ($T_\text{dead}$) reflects the combined effects of the coil ring-down (with time-constant $\tau$) and the saturation of the RF line following the Tx pulse. If acquisition begins while the energy stored in the resonator still competes with the amplitude of the NMR signal, the low-frequency samples of $k$-space will contain a spurious contribution that often manifests as a sinc-type artifact in the image (see Fig.~\ref{fig:optimizacionpetra}a)I). Furthermore, the duration of the acquisition window ($T_\text{acq}$) is constrained by the coherence time of the NMR signal ($T_2^*$) evolving under $B_0(\vec{r})$. When the acquisition extends beyond this limit (i.e., $T_\text{acq}>T_2^*$), the reconstruction exhibits apodization and blurring artifacts at the image periphery (see Fig.~\ref{fig:optimizacionpetra}a)V). Both factors are particularly unfavorable in our regime. On one hand, $\tau$ scales as $\propto 2Q/\omega_\text{L}$, where $Q$ is the quality factor of the coil and $\omega_\text{L}$ is the Larmor frequency, so ring-down times are particularly long for high-$Q$ coils in LF-MRI. On the other, the lightweight and compact design of the Halbach magnet compromises $B_0$ homogeneity, limiting the maximum achievable transverse coherence time to $T_2^*\sim1$\,ms even in substances with $T_2>1$\,s (see Fig.~\ref{fig:calibraciones}b).

\subsection{RF pulse quality}
\label{sec:pulse}

In any ZTE sequence, spin excitation occurs once the encoding gradient has reached its set value. To ensure uniform and coherent excitation of all spins within the sample, very short rectangular RF pulses are essential \cite{Weiger2011}. Only in this way does the excitation bandwidth uniformly cover the spectral range given by $\Delta \omega = \gamma \cdot (\Delta \vec{B}_0(\vec{r}) + |\vec{g}_\text{enc}(\vec{r},t)|)$, with $\vec{g}_\text{enc}$ the gradient field for spatial encoding and $\gamma$ the $g$-factor ($\gamma=2\pi \cdot42.57$\,MHz/T for $^{1}\text{H}$). Achieving relevant flip angles (FAs) with such short pulses, however, demands high-power RF amplification. For our purposes, we prefer a high-$Q$ coil (to maximize SNR) and RF pulses with a square envelope and maximum duration of $t_\text{RF}=30$\,µs (see Fig.~\ref{fig:optimizacionpetra}c). In this regime, $B_1$ pulses can be suboptimal, since the expected rectangular profile of $\vec{B}_1(t)$ is dominated by the current rise time (see Fig.~\ref{fig:preemphasis}a). This distortion acts as a low-pass filter, reducing the excitation bandwidth and causing FA degradation toward the peripheral regions of the sample.

\subsection{System heating}
\label{sec:heat}

Another challenge is the requirement for high continuous gradient currents during each repetition. While this minimizes eddy currents, it may compromise thermal stability during long studies, particularly for sequences with $\text{TR} \gg T_1$. Scanners designed for ZTE typically employ water-cooled gradient systems \cite{Algarin2020}, but this is impractical for portable designs. In our case, the high efficiency and low resistance of the copper plate-based gradients keep Joule effects low (see Sec.~\ref{sec:hardware}), so 2\,mm-resolution PETRA sequences lead to minor heating even for $\text{TR}=250$\,ms ($g_\text{enc}=5.8$\,mT/m, $I_x=9.5$\,A, so $P<8$\,W in the worst-case scenario and assuming full duty cycle). Nevertheless, during long \emph{in-vivo} acquisitions exceeding one hour, the Larmor frequency may drift (see Fig.~\ref{fig:calibraciones}d).

\subsection{Sequence duration}
\label{sec:duration}

A further drawback of PETRA sequences is their relatively long acquisition time, primarily due to: (i) the need to sample the pointwise region of $k$-space; (ii) the impossibility of forming long echo trains to fill multiple $k$-space lines per repetition; and (iii) the inherently three-dimensional nature of ZTE. Although sequences combining slice selection with ZTE are available \cite{Borreguero2024}, such approaches demand high homogeneity in all magnetic fields, which is difficult to achieve in portable Halbach systems. The duration of a PETRA sequence depends on the matrix size and maximum field of view ($N_\text{max}$ and $\text{FoV}_\text{max}$), the radial undersampling factor ($\text{US}_\text{rad}$), and the ratio $T_\text{dead}/T_\text{acq}$, according to:
\begin{equation}
	T_{\text{seq}} \; \propto \; N_\text{av} \cdot \text{TR} \left[ \frac{\pi}{2} \cdot \frac{N_\text{max}^2}{\text{US}_\text{rad}}  + \frac{4}{3} \pi \left( \frac{N_\text{max}}{2 \cdot \text{FoV}_\text{max}} \frac{T_\text{dead}}{T_\text{acq}} \right)^3 \right],
\end{equation}
where $N_\text{av}$ is the number of averages. For typical knee imaging resolutions of 2\,mm with fully sampled $k$-space, a ratio of $T_\text{dead}/T_\text{acq}=200$\,µs/1\,ms implies that approximately 15\,\% of the total sequence time is spent filling the central region of $k$-space point by point ($N_\text{rad}=25,250$, $N_\text{sp}=3,616$), making PETRA relatively time-inefficient.

\subsection{Imaging in inhomogeneous $B_0$ and $B_1$ fields}

ZTE sequences are particularly sensitive to artifacts arising from $B_0$ inhomogeneity, due to their radial acquisition of signals undergoing $T_2^*$ decay without spin refocusing. Under these circumstances, the reconstruction of $k$-space using algorithms agnostic to the physics of the problem tends to exhibit blurring, anisotropic degradation of resolution, and hyperintense bands in pixels where the encoding field ($\vec{B}_0 (\vec{r}) + \vec{g}_\text{enc}(\vec{r},t)$) is extremal. In addition, the use of solenoidal coils results in a spatial modulation of both the transmission and reception efficiencies. This fact causes two effects: (i) a defective FA calibration, since the Rabi flops reflect the integrated contribution of all spins that have been excited with a position-dependent FA, which depends on the deviation of the $B_1$ field from its average value; and (ii) a spatial modulation of brightness in the reconstruction even if the FA is homogeneous across the entire sample. This indeterminacy is especially critical when employing a variable-flip-angle (VFA) method for $T_1$ quantification (see Sec.~\ref{sec:theory_relax_maps}). To overcome the uncertainties caused by the presence of magnetic field inhomogeneities, prior knowledge of the distributions of $B_0(\vec{r})$ and $B_1(\vec{r})$ is essential.

\subsection{Mapping the $B_0$ and $B_1$ fields}
\label{sec:theory_maps}

\begin{figure*}
	\includegraphics[width=1.0\textwidth]{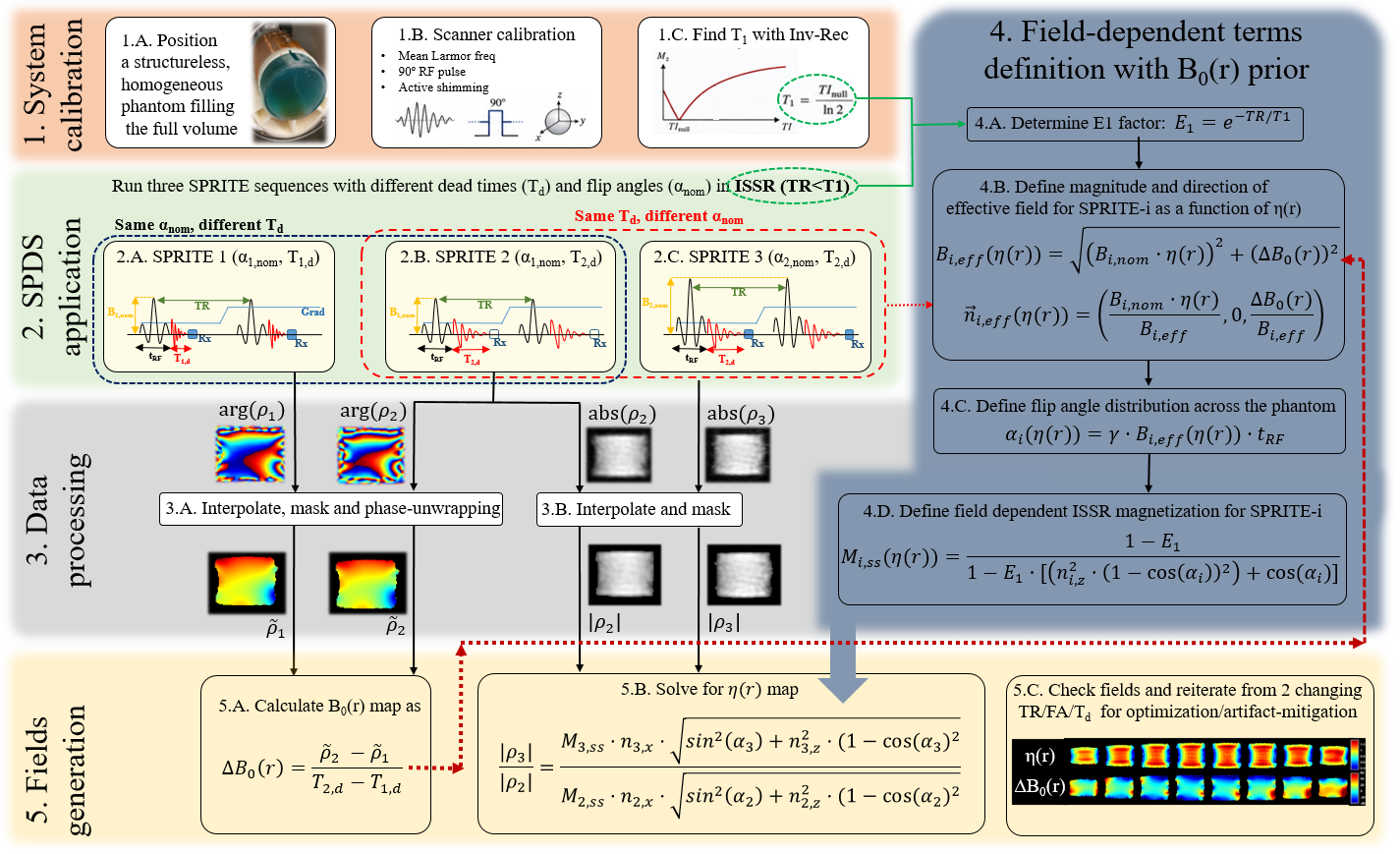}
	\caption{Extended Single-Point Double-Shot (SPDS) pipeline for simultaneous $B_0(\vec{r})$ and $B_1(\vec{r})$-efficency mapping.}
	\label{fig:spdspipeline}
\end{figure*}

For $B_0$ and $B_1$ mapping, one can adopt the strategy proposed in Ref.~\cite{Borreguero2024}, extending its use to simultaneously obtain maps of $B_0(\vec{r})$ and $B_1(\vec{r})$ with immunity to spatial distortions. A schematic of the pipeline is shown in Figure~\ref{fig:spdspipeline}. The extended method is based on three fast, low-resolution SPRITE sequences that are identical except for their nominal flip angles ($\alpha_\text{\emph{i},nom}$) and dead times ($T_\text{\emph{i},d}$), each producing a reconstruction $\rho_i$. Ideally, these sequences should be applied to a structureless phantom with homogeneous and known $T_1$ that fills the available imaging volume.

Analogously to the SPDS approach, the phase maps $\text{arg}(\rho_{1})$ and $\text{arg}(\rho_{2})$ are cropped to a common mask and subsequently unwrapped in three dimensions. Their subtraction yields a map of deviations in the main field:
\begin{equation}
	\Delta B_{0} (\vec{r}) \approx
	\frac{\text{arg}(\rho_{2})-\text{arg}(\rho_{1})}{
		\gamma \left( T_\text{2,d}-T_\text{1,d} \right)}.
	\label{eq:B0fromGE}
\end{equation}

The magnitude maps $\text{abs}(\rho_{1})$ and $\text{abs}(\rho_{3})$ exhibit a common spatial modulation of brightness, such that the local flip angle $\alpha_i(\vec{r})$ deviates from its nominal value, estimated as $\alpha_\text{\emph{i},nom} = \gamma \, B_{1_\text{\emph{i},nom}} \, t_\text{RF}$ (assuming negligible $B_0$ inhomogeneity). The actual FA distribution is therefore:
\begin{equation}
	\alpha_{i}(\vec{r}) = \gamma \, B_\text{\emph{i},eff}(\vec{r}) \, t_\text{RF},
	\label{eq:alphanominal}
\end{equation}
where $B_\text{\emph{i},eff}(\vec{r}) = \sqrt{(B_{1_\text{\emph{i},nom}} \, \eta (\vec{r}))^2 + (\Delta B_0 (\vec{r}))^2}$ is the position-dependent effective field in the rotating frame, and $\eta (\vec{r})$ is a dimensionless map representing the scaling between the nominal $B_1$ strength ($B_{1_\text{\emph{i},nom}}$, calibrated from Rabi flops and corresponding to the spatially averaged $B_1$ amplitude within the sample) and the actual $B_1$ at each pixel within the region of interest (RoI):
\begin{equation}
	\label{eq:eta}
	B_{1}(\vec{r}) = B_\text{1,nom} \cdot \eta(\vec{r}).
\end{equation}

Because $T_1$ is known a priori and homogeneous throughout the imaging volume, the relative brightness modulation between SPRITE images with different flip angles arises solely from $B_1$ inhomogeneity. Therefore, the pixelwise ratio of $\text{abs}(\rho_{1})$ and $\text{abs}(\rho_{3})$ is given by the ratio of their steady-state signals:
\begin{equation}
	\label{eq:facalibration}
	\frac{\text{abs}(\rho_1)}{\text{abs}(\rho_3)} =
	\frac{
		M_\text{1,ss} \, n_{1,x} \, \sqrt{ \sin^2(\alpha_1) + n_{1,z}^2 \, \bigl(1 - \cos(\alpha_1)\bigr)^2 }}
	{M_\text{3,ss} \, n_{3,x} \, \sqrt{ \sin^2(\alpha_3) + n_{3,z}^2 \, \bigl(1 - \cos(\alpha_3)\bigr)^2 }}.
\end{equation}
Here, $n_{i,x}$ and $n_{i,z}$ denote the components of the unit vector pointing along $\vec{B}_\text{eff}$ in the rotating frame:
\begin{equation}
	\label{eq:vec_n}
	\vec{n}_i (\vec{r}) = 
	\left( 
	\frac{B_{1_\text{\emph{i},nom}} \, \eta (\vec{r})}{B_{\text{\emph{i},eff}} (\vec{r})}, 0, 
	\frac{\Delta B_0 (\vec{r})}{B_{\text{\emph{i},eff}} (\vec{r})} 
	\right),
\end{equation}
and $M_\text{\emph{i},ss}$ is the longitudinal magnetization reached in the incoherent steady state:
\begin{equation}
	\label{eq:Mss}
	M_\text{\emph{i},ss} = 
	\frac{1 - E_1}{
		1 - E_1 \Bigl( n_{i,z}^2 \bigl(1 - \cos(\alpha_i)\bigr) + \cos(\alpha_i) \Bigr)
	},
\end{equation}
where $E_1 = \exp(-\text{TR}/T_1)$.

Once the matrix $\text{abs}(\rho_{1}) / \text{abs}(\rho_{3})$ is computed and masked to select the phantom region, Eq.~(\ref{eq:facalibration}) can be solved for $\eta(\vec{r})$ at each voxel, using the $\Delta B_0(\vec{r})$ map obtained from Eq.~(\ref{eq:B0fromGE}). This can be used for mapping the effective $B_1(\vec{r})$ field strength according to Eq.~(\ref{eq:eta}).

This method effectively decouples the combined effects of $B_0$ and $B_1$ inhomogeneities on their respective maps. On one hand, $B_1$ inhomogeneity introduces a common brightness modulation across the three images, but this does not affect the $B_0$ map, since its estimation relies solely on the phase of $\rho$, which is insensitive to amplitude variations. On the other, $B_0$ inhomogeneity induces geometric distortions that could otherwise deform the maps. However, the use of SPRITE sequences with encoding times below 500\,µs ensures robustness of the $\rho$ images against such distortions, even when reconstructed using general algorithms such as the FFT. Moreover, since the pair of images used to define the $B_1$ field share identical encoding times, their geometric deformation is the same and could, if necessary, be corrected by reconstructing them using the previously obtained $B_0$ map as a prior. As a result, the ratio of their magnitudes remains exact and free from spatial mixing effects due to $B_0$ inhomogeneity.

\subsection{Standard relaxation mapping}
\label{sec:theory_relax_maps_std}

The literature on the relaxation times of human tissues at low magnetic fields remains scarce, and models predicting their field dependence have been only marginally validated with \emph{in-vivo} data \cite{Bottomley1984,Korb2002,Rooney2007}. Although some reports of $T_1$ and $T_2$ measurements exist for $B_0$ values close to ours \cite{OReilly2021,Sarracanie2021}, we chose to independently measure the relaxation parameters of fat and muscle in the knee at $B_0=90$\,mT. For this purpose, we employed classical quantification methods for $T_2$ and $T_1$ based on images acquired with Cartesian multi–spin-echo sequences and variable TE/TI, as described in Sec.~\ref{sec:goldstandardmethods}. More efficient approaches such as MR fingerprinting \cite{Ma2013} were not used, as they proved impractical in our control system due to current data streaming limitations. This initial strategy for quantifying $T_1$ and $T_2$ serves as a reference for the expected relaxation times of fat and muscle, which will later be used for comparison with values obtained using more advanced methods such as VFA-PETRA.

\subsection{VFA-PETRA relaxation mapping}
\label{sec:theory_relax_maps}

$T_1$ mapping using the variable flip angle (VFA) method is a long-established approach \cite{FRAM1987201,GUPTA1977231} widely used in high-field MRI for rapid relaxometry of soft tissues. Recently, the VFA method has been successfully combined with sequences designed for short-$T_2$ components, such as UTE for mouse lungs \cite{Alamidi} and ZTE for the human brain \cite{Ljungberg2020}. However, these reports were obtained in high-performance clinical systems operating at $B_0 \geq 3$\,T, where SNR and field inhomogeneities are not critical limitations. 

When $B_1$ is inhomogeneous but $B_0$ remains highly uniform, \emph{in-vivo} $T_1$ mapping of soft tissues can be achieved by combining VFA with spoiled gradient-echo (SPGR) sequences, provided that accurate $B_1$ information is incorporated into the model \cite{Liberman2014,Lee2017}. This method can be extended to very low-field systems, as recently demonstrated in Ref.~\cite{Shen2025} at 6.5\,mT for \emph{in-vivo} breast $T_1$ mapping, although the SNR drop limits the precision of the method.

To enable simultaneous $T_1$ mapping of hard and soft tissues, this approach can be extended by replacing SPGR with a pair of PETRA sequences with $\text{TR} \ll T_1$ and different nominal flip angles ($\alpha_\text{nom}$), hereafter referred to as VFA-PETRA. Because PETRA maintains the gradients at a constant level after the receive window, they act as effective spoilers of the transverse magnetization, ensuring the establishment of an incoherent steady state across all tissues. To minimize artifacts inherent to this regime, several dummy pulses can be applied before the repetitions in which data were acquired.

Unlike the SPDS strategy, in which the known and spatially homogeneous $T_1$ of a phantom is used to estimate $B_0(\vec{r})$ and $B_1(\vec{r})$ variations, VFA-PETRA exploits the fact that $B_0(\vec{r})$ and $B_1(\vec{r})$ are already known from a prior SPDS experiment. In this case, $T_1(\vec{r})$ becomes the only spatially-varying unknown. 

With the $\rho_{\alpha_1}$ and $\rho_{\alpha_2}$ reconstructions using PK of $B_0$, the noise background can be removed using a common mask applied to both magnitude images. For all voxels within the mask, the ratio 
\[
\lambda(\vec{r}) = \frac{\text{abs}(\rho_{\alpha_1})}{\text{abs}(\rho_{\alpha_2})}
\]
can be computed and solved for $T_1(\vec{r})$ according to the model:
\begin{equation}
	\lambda(\vec{r}) =
	\frac{
		M_\text{1,ss}(T_1(\vec{r})) \, n_{1,x} \, 
		\sqrt{ \sin^2(\alpha_1) + n_{1,z}^2 \, \bigl(1 - \cos(\alpha_1)\bigr)^2 }
	}{
		M_\text{2,ss}(T_1(\vec{r})) \, n_{2,x} \, 
		\sqrt{ \sin^2(\alpha_2) + n_{2,z}^2 \, \bigl(1 - \cos(\alpha_2)\bigr)^2 }
	},
	\label{eq:lambdaVFAratio}
\end{equation}
where $\alpha_{i}$, $\vec{n}_i$, and $M_\text{\emph{i},ss}$ are defined by Eqs.~(\ref{eq:alphanominal}), (\ref{eq:vec_n}), and (\ref{eq:Mss}), respectively, each including its voxel dependence through $B_\text{\emph{i},eff}\bigl(B_1(\vec{r}), B_0(\vec{r})\bigr)$, as determined from a previous SPDS experiment.


\section{Methods}
\label{sec:methods}

\subsection{Scanner}
\label{sec:hardware}

\begin{figure*}
	\includegraphics[width=1\textwidth]{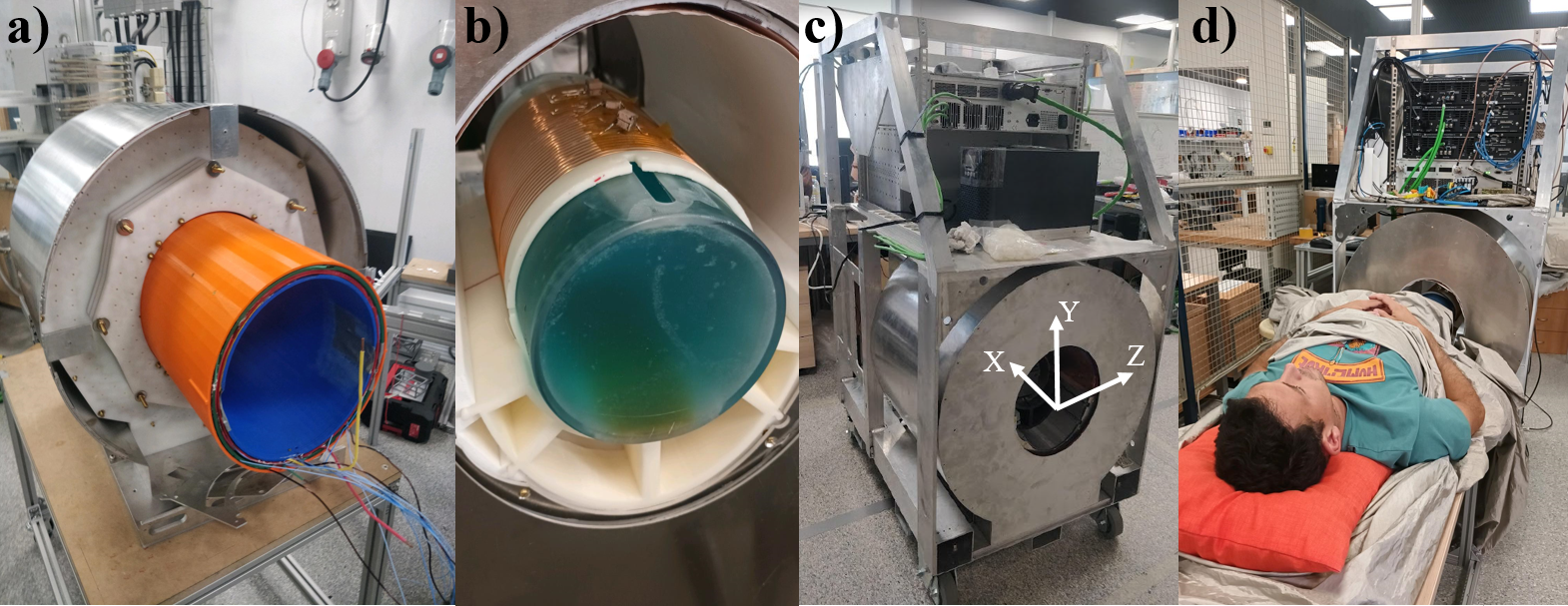}
	\caption{The scanner used for the experiments here presented. a) Elliptical Halbach magnet after ring joining and gradients module inserted. b) RF solenoidal coil loaded with a phantom for $B_0$ and $B_1$ mapping of full volume. c) General outlook of the MRI system in the laboratory once finished the integration of all componentes. d) Patient positioning during a knee study in this system.}
	\label{fig:fig_scanner}
\end{figure*}

All results have been obtained on a portable 90~mT MRI system ($\approx 3.64$~MHz for $^{1}\text{H}$) designed for extremity and brain imaging \cite{galve2024elliptical}.

The magnet is based on an elliptical Halbach arrangement constructed from Nd$_2$Fe$_{14}$B cubes (Fig.~\ref{fig:fig_scanner}a) and incorporates a shimming unit, resulting in a main field of $\approx90$\,mT with approximately 4,000\,ppm over a sphere of 20\,cm in diameter. The compact design of the magnet (the bore has minor and major axes of $d_\text{m}=20$\,cm and $d_\text{M}=28$\,cm, and a total axial length of $l=44$\,cm) provides a high degree of portability while still offering sufficient space to position the head within the FoV, despite anatomical constraints imposed by the shoulders.

The gradient coils consist of three modules, each comprising four concentric elliptical lobes manufactured from water-jetted copper plates (Fig.~\ref{fig:fig_scanner}a). The gradient coils have efficiencies of (0.61, 0.87, 0.87)\,mT/m/A and resistances of (88, 128, 112)\,m$\Omega$. The coils are driven by three AE Techron 7224 power amplifiers (Indiana, USA), which can deliver currents up to 50\,A to the loads, achieving maximum gradient strengths of (30.5, 43.5, 43.5)\,mT/m.

The RF system is based on a single solenoidal coil 15\,cm in length and diameter, wound with 44 turns of 2\,mm copper wire on a PLA holder (see Fig.~\ref{fig:fig_scanner}b). The coil bandwidth was $\approx260$\,kHz when loaded with a knee, corresponding to a quality factor of $Q\approx250$\footnote{The quality factor ($Q$) was determined from the time constant of the ring-down signal ($\tau$) following an RF pulse generated with a square input to the RFPA. The magnetic field $B_1(t)$ was recorded on an oscilloscope via flux induction in a pickup loop near the RF coil (see Fig.~\ref{fig:preemphasis}a), and the decay envelope was fitted to an exponential model.} ($\tau\approx22$\,µs; see Fig.~\ref{fig:preemphasis}a). The RF low-noise amplifier (45\,dB gain, noise factor $\approx$1\,dB) and power amplifier (250\,W, maximum duty cycle 10\,\% with 10\,ms pulses), as well as the passive Tx/Rx switch, were purchased from Barthel HF-Technik GmbH (Aachen, Germany).

All components are housed within an aluminum structure (Fig.~\ref{fig:fig_scanner}c) measuring $68\times95\times140$\,cm$^3$ (width $\times$ height $\times$ length) and weighing 377\,kg in total, allowing the system to be easily moved. The device operates as a standalone unit and requires only a connection to a standard single-phase power outlet.

During knee imaging experiments, the body acts as an antenna \cite{Kibret}, coupling significant amounts of electromagnetic interference into the RF coil. To minimize this effect, the coil and subject are partially covered with conductive cloth from Holland Shielding Systems BV (Dordrecht, the Netherlands) (Fig.~\ref{fig:fig_scanner}d). All images were acquired with an RMS noise floor below $<1.5\times$ the expected thermal noise limit \cite{guallart2025electromagnetic}, with the RF coil impedance matched to $50.0\pm0.2$\,$\Omega$ at the daily operating frequency.

The scanner is controlled through the open-source MaRCoS (Magnetic Resonance Control System) framework \cite{Guallart2022,NEGNEVITSKY2023107424}, which features synchronized pulse-sequence execution and data acquisition. User interaction with MaRCoS is handled via the MaRGE interface (MaRCoS Graphical Environment) \cite{ALGARIN2024107662}, and all pulse sequences used in this work are openly available on GitHub \cite{githubmarge}.

\subsection{Pulse sequences}
\label{sec:sequences}

\begin{table*}[t]
	\centering
	\caption{Acquisition parameters for PETRA sequences.}
	\label{tab:PETRAimageParameters}
	\resizebox{\textwidth}{!}{
		\begin{tabular}{|c|c|c|c|c|c|c|c|c|c|c|c|c|c|}
			\hline
			Image  & \begin{tabular}[c]{@{}c@{}} FA ($^\circ$)\end{tabular} & \begin{tabular}[c]{@{}c@{}}RF pulse \\ length (µs)\end{tabular} & \begin{tabular}[c]{@{}c@{}}FoV\\ (cm$^3$)\end{tabular} & \begin{tabular}[c]{@{}c@{}}Pixel \\ size\\ (mm)\end{tabular} & \begin{tabular}[c]{@{}c@{}}Dead\\ time (µs) /\\ Acquisition\\ time (ms)\end{tabular} & \begin{tabular}[c]{@{}c@{}}Bandwidth\\ (kHz)\end{tabular}  & \begin{tabular}[c]{@{}c@{}}TR\\ (ms)\end{tabular} & \begin{tabular}[c]{@{}c@{}} US$_{rad}$ \end{tabular} & \begin{tabular}[c]{@{}c@{}}Radial\\ spokes \end{tabular} & \begin{tabular}[c]{@{}c@{}}Pointwise\\ spokes\end{tabular} & Averages & \begin{tabular}[c]{@{}c@{}}Scan\\ time\\ (min)\end{tabular} \\ \hline
			
			Fig.~\ref{fig:optimizacionpetra}a.I  & 90   & 4/20/3   & 20$\times$16$\times$16 & 2$\times$2$\times$2 & 150 / 1               & 50  & 50  & 1 & 25,250   & 568  & 1    & 21.5  \\ \hline

			Fig.~\ref{fig:optimizacionpetra}a.II  & 90   & 4/20/3   & 20$\times$16$\times$16 & 2$\times$2$\times$2 & 175 / 1               & 50  & 50 & 1  & 25,250   & 1,584  & 1    & 22.4 \\ \hline
			
			Fig.~\ref{fig:optimizacionpetra}a.III  & 90   & 4/20/3   & 20$\times$16$\times$16 & 2$\times$2$\times$2 & 200 / 1               & 50  & 50 & 1  & 25,250   & 3,416  & 1    & 23.9  \\ \hline

			Fig.~\ref{fig:optimizacionpetra}a.IV & 90   & 4/20/3   & 20$\times$16$\times$16 & 2$\times$2$\times$2 & 200 / 2               & 50  & 50 & 1  & 25,250   & 488  & 1    & 21.4  \\ \hline
			
			Fig.~\ref{fig:optimizacionpetra}a.V  & 90   & 4/20/3   & 20$\times$16$\times$16 & 2$\times$2$\times$2 & 200 / 3               & 50  & 50 & 1  & 25,250   & 168  & 1    & 21.2  \\ \hline

			
			Fig.~\ref{fig:optimizacionpetra}b.I  & 90   & 4/20/3   & 20$\times$16$\times$16 & 2$\times$2$\times$2 & 200 / 1               & 50  & 50 & 1  & 25,250   & 3,416  & 1    & 23.9  \\ \hline
			
			Fig.~\ref{fig:optimizacionpetra}b.II  & 90   & 4/20/3   & 20$\times$16$\times$16 & 2$\times$2$\times$2 & 200 / 1               & 50  & 50 & 2  & 12,622   & 3,416  & 1    & 13.4  \\ \hline
			
			Fig.~\ref{fig:optimizacionpetra}b.III  & 90   & 4/20/3   & 20$\times$16$\times$16 & 2$\times$2$\times$2 & 200 / 1               & 50  & 50 & 4  & 6,370   & 3,416  & 1    & 8.2  \\ \hline
			
			Fig.~\ref{fig:optimizacionpetra}b.IV  & 90   & 4/20/3   & 20$\times$16$\times$16 & 2$\times$2$\times$2 & 200 / 1               & 50  & 50 & 8  & 3,226   & 3,416  & 1    & 5.5  \\ \hline
			
			Fig.~\ref{fig:optimizacionpetra}b.V  & 90   & 4/20/3   & 20$\times$16$\times$16 & 2$\times$2$\times$2 & 200 / 1               & 50  & 50 & 16  & 1,604   & 3,416  & 1    & 4.2  \\ \hline

			
			Fig.~\ref{fig:optimizacionpetra}c.I  & 90   & 4/10/3   & 20$\times$16$\times$16 & 2$\times$2$\times$2 & 200 / 1               & 50  & 50 & 1  & 25,250   & 3,192  & 1    & 23.7  \\ \hline
			
			Fig.~\ref{fig:optimizacionpetra}c.II  & 90   & 4/20/3   & 20$\times$16$\times$16 & 2$\times$2$\times$2 & 200 / 1               & 50  & 50 & 1  & 25,250   & 3,416  & 1    & 23.8  \\ \hline
			
			Fig.~\ref{fig:optimizacionpetra}c.III  & 90   & 4/50/3   & 20$\times$16$\times$16 & 2$\times$2$\times$2 & 200 / 1               & 50  & 50 & 1  & 25,250   & 4,176  & 1    & 24.5  \\ \hline
			
			Fig.~\ref{fig:optimizacionpetra}c.IV  & 90   & 4/150/3   & 20$\times$16$\times$16 & 2$\times$2$\times$2 & 200 / 1               & 50  & 50 & 1  & 25,250   & 7,512  & 1    & 27.3  \\ \hline
			
			Fig.~\ref{fig:optimizacionpetra}c.V  & 90   & 4/300/3   & 20$\times$16$\times$16 & 2$\times$2$\times$2 & 200 / 1               & 50  & 50 & 1  & 25,250   & 15,096  & 1    & 33.62  \\ \hline
			

			Fig.~\ref{fig:fieldsmapping}c\&d  & 90   & 4/20/3   & 20$\times$16$\times$16 & 2$\times$2$\times$2 & 200 / 1               & 40  & 50 & 1  & 25,250   & 1,936  & 1    & 22.6  \\ \hline
			
			
			Fig.~\ref{fig:calibracionVAFPETRA}a & 10:10:90   & 4/20/3   & 20$\times$16$\times$16 & 2$\times$2$\times$2 & 300 / 1               & 50  & 10 & 5  & 5,108   & 10,600  & 2    & 5.2  \\ \hline
			
			
			Fig.~\ref{fig:kneeisotropica} top  & 40   & 2.5/20/2.5   & 20$\times$16$\times$16 & 1.6$\times$1.6$\times$1.6 & 300 / 1.25               & 48  & 20 & 4  & 9,488   & 10,752  & 2    & 13.5  \\ \hline
			
			Fig.~\ref{fig:kneeisotropica} bottom  & 30   & 2.5/10/2.5   & 20$\times$16$\times$12 & 1.6$\times$1.6$\times$1.6 & 320 / 1.25               & 48  & 25 & 4  & 7,086   & 9,096  & 2    & 13.4  \\ \hline

			
			Fig.~\ref{fig:rarevspetra}a\&b.IV  & 30   & 2.5/20/2.5   & 20$\times$16$\times$16 & 2$\times$2$\times$4 & 200 / 1               & 50  & 50 & 2  & 6,380   & 1,864  & 1    & 6.8  \\ \hline
			
			Fig.~\ref{fig:rarevspetra}c\&d.IV  & 40   & 2.5/20/2.5   & 20$\times$16$\times$12 & 2$\times$2$\times$3 & 300 / 1               & 50  & 20 & 4  & 3,232   & 5,151  & 3    & 8.4  \\ \hline
			
			
			Fig.~\ref{fig:FAandTRsweep}  & 20:20:80   & 4/20/2.5   & 20$\times$16$\times$16 & 2$\times$2$\times$4 & 200 / 1.0               & 48  & 25/50/100 & 4  & 3,232   & 1,832  & 4/2/1    & 8.4  \\ \hline
			

			Fig.~\ref{fig:contrast}\&\ref{fig:mapasrodillapetra}a  & 90   & 4/20/2.5   & 20$\times$16$\times$16 & 2$\times$2$\times$4 & 200 / 1.25               & 48  & 50 & 4  & 3,842   & 3,568  & 1    & 6.1  \\ \hline
			
			
			Fig.~\ref{fig:brainpetra}  & 30   & 0/25/0   & 24$\times$24$\times$24 & 3$\times$3$\times$3 & 300 / 1.5               & 26  & 100 & 5  & 4,088   & 2,800  & 1    & 11.4  \\ \hline

	\end{tabular}}
\end{table*}

\begin{table*}[t]
	\centering
	\caption{Acquisition parameters for RARE sequences.}
	\label{tab:RAREimageParameters}
	\resizebox{\textwidth}{!}{
		\begin{tabular}{|c|c|c|c|c|c|c|c|c|c|c|c|c|c|}
			\hline
			Image  & \begin{tabular}[c]{@{}c@{}}Pulse \\ time (µs)\end{tabular} & \begin{tabular}[c]{@{}c@{}}FoV\\ (mm$^3$)\end{tabular} & \begin{tabular}[c]{@{}c@{}}Pixel \\ size\\ (mm)\end{tabular} & \begin{tabular}[c]{@{}c@{}}TI (ms)\end{tabular} & \begin{tabular}[c]{@{}c@{}}TE (ms)\end{tabular} & \begin{tabular}[c]{@{}c@{}} ETL \end{tabular} &  \begin{tabular}[c]{@{}c@{}}TR (ms)\end{tabular} &   \begin{tabular}[c]{@{}c@{}}$T_{acq}$ (ms)\end{tabular} &  \begin{tabular}[c]{@{}c@{}}Dummy \\ pulses\end{tabular} & \begin{tabular}[c]{@{}c@{}}Partial\\Fourier (\%) \end{tabular} & Averages & \begin{tabular}[c]{@{}c@{}}Scan\\ time\\ (min)\end{tabular} \\ \hline
			
			Fig.~\ref{fig:mapasrodillaclasicos}a  & 28/56   & 20$\times$16$\times$16 & 2$\times$2$\times$4 & 0 & 20:100  & 4 & 800 & 4  & 0  & 70 & 1    & 7.4  \\ \hline
			
			Fig.~\ref{fig:mapasrodillaclasicos}b  & 32/64   & 20$\times$16$\times$16 & 2$\times$2$\times$4 & 0:600 & 20  & 4 & 800 & 4  & 0  & 70 & 1    & 7.4  \\ \hline
			Fig.~\ref{fig:rarevspetra}a-d.I  & 100   & 20$\times$16$\times$16 & 2$\times$2$\times$4 & 0 & 15  & 4 & 800 & 2  & 0  & 70 & 1    & 7.5  \\ \hline
			
			Fig.~\ref{fig:rarevspetra}a-d.II & 100   & 20$\times$16$\times$16 & 2$\times$2$\times$4 & 0 & 40  & 8 & 800 & 2  & 0  & 70 & 2    & 7.5  \\ \hline
			
			Fig.~\ref{fig:rarevspetra}a-d.III  & 100   & 20$\times$16$\times$16 & 2$\times$2$\times$4 & 0 & 20  & 4 & 120 & 2  & 10  & 70 & 7    & 7.8  \\ \hline

	\end{tabular}}
\end{table*}

The pulse sequence parameters for all qualitative and quantitative acquisitions are provided in Table~\ref{tab:PETRAimageParameters} (PETRA) and Table~\ref{tab:RAREimageParameters} (RARE).

\subsection{System calibrations and PETRA optimization}
\label{sec:met_calibration}

To alleviate the effects of coil ringing and Tx/Rx switching delays, we adopted the pre-emphasis and counter-emphasis approach proposed in Ref.~\cite{Ha2022}. In this scheme, the RF pulse is composed of three consecutive components: (0) a short, high-amplitude pre-emphasis pulse of the same phase as the main excitation; (1) the main excitation pulse; and (2) a counter-emphasis pulse of opposite phase (180° phase shift) that suppresses residual energy stored in the resonator (see Fig.~\ref{fig:preemphasis}b). This strategy effectively reduces post-transmission ringing\footnote{Not to be confused with a reduction in the intrinsic ring-down time constant, which remains unchanged.}. 

The impact of pre/counter-emphasis on RF pulse quality (Sec.~\ref{sec:pulse}) was evaluated by recording the $B_1(t)$ field trace on an oscilloscope using a small pickup coil (2\,cm diameter, 4 turns) positioned close to and approximately concentric with the Tx/Rx coil. The voltage and duration configuration used was $V_\text{RF,0}/V_\text{RF,1}/V_\text{RF,2}=1/0.25/1$\,a.u. and $t_\text{RF,0}/t_\text{RF,1}/t_\text{RF,2}=4/50/3$\,µs, corresponding respectively to the pre-, main-, and counter-emphasis pulses. In the configuration without pre-emphasis, only the main component was applied with $V_\text{RF,1}=0.25$\,a.u. and $t_\text{RF,1}=50$\,µs.

To prolong $T_2^*$ times (Sec.~\ref{sec:times}), we shimmed actively prior to each study, based on the line-narrowing of the free NMR signal spectrum (see Fig.~\ref{fig:calibraciones}c). For this step, the pre-emphasis configuration was $V_\text{RF,0}/V_\text{RF,1}/V_\text{RF,2}=1/0.25/1$\,a.u. and $t_\text{RF,0}/t_\text{RF,1}/t_\text{RF,2}=3/25/3$\,µs.

To correct for slow Larmor-frequency drifts (Sec.~\ref{sec:heat}), the carrier and demodulation frequencies were recalibrated at intervals of no more than 10\,minutes.

To keep acquisition times below 15\,min (Sec.~\ref{sec:duration}), we employed a moderate degree of radial undersampling (US$_\text{rad}$ up to $\times4$). However, the pointwise region of $k$-space is always fully acquired, as compressed sensing \cite{Ilbey2022,FABICH2014116} was suboptimal for our context. Although the gradient system would tolerate longer TRs without overheating, we limited the maximum TR to 50\,ms.

\subsection{Field maps}
\label{sec:fieldmapping}

We followed the procedure introduced in Sec.~\ref{sec:theory_maps} and depicted in Figure~\ref{fig:spdspipeline} to obtain simultaneous $B_0(\vec{r})$ and $B_1(\vec{r})$ maps in our 90\,mT system, where the Halbach magnet and the single, uniformly wound solenoidal RF coil introduce significant fluctuations in both fields. The three SPRITE sequences were applied to a structureless phantom filled with a 1\,\% CuSO$_4$ solution ($\text{TI}=8.6$\,ms, $T_1=12.5$\,ms, measured by inversion recovery) that fully occupied the RF imaging volume (see Fig.~\ref{fig:fig_scanner}b). 

Common parameters for the SPRITE sequences were: $N_\text{av}=1$, $\text{FoV}=(20,16,16)$\,cm, $N=(20,16,16)$, $\Delta r=10\times10\times10$\,mm$^3$, $T_\text{1,d}/T_\text{2,d}=350/550$\,µs, $\alpha_\text{1,nom}/\alpha_\text{2,nom}=60/80$°, $\text{TR}=10$\,ms, $\text{BW}=50$\,kHz, $N_\text{sp}=10{,}000$, $N_\text{dummy}=10$, and $T_\text{seq}=100$\,s, with a total protocol duration of $T_\text{prot}=5$\,min. 

For the first and third sequences, the different flip angles were generated by increasing the nominal $B_1$ strength while keeping the RF pulse length ($t_\text{RF}=20$\,µs) and pre/counter-emphasis identical in both cases, thereby avoiding relative modulations due to differing excitation bandwidths. The three acquired $k$-spaces were zero-padded by a factor of five to increase the spatial resolution to 2\,mm and subsequently reconstructed using FFT to obtain the complex matrices $\rho_i$.

To estimate the spatially varying $B_1$-scaling, we solved Eq.~\ref{eq:facalibration} for the value of $\eta$ that minimizes the $L_2$ norm of the discrepancy between the analytical steady-state signal-ratio model (right-hand side of Eq.~\ref{eq:facalibration}) and the experimentally measured ratio at each voxel (left-hand side), using in MATLAB a derivative-free Nelder–Mead search (\texttt{fminsearch}) initialized at $\eta$=1 for all pixels. Finally the raw arrays representing $B_0(r)$ and $B_1(r)$ in the FoV were denoised using a Gaussian filter (\texttt{imgaussfilt} function in MATLAB with $\sigma$=1.5); otherwise, noise in the maps could propagate to the final image.

Note that the SPDS procedure was executed only once prior to the experiments with a time cost $<$5~min, and the resulting $B_0$ and $B_1$ field models were used to reconstruct all subsequent datasets acquired over several days. Overall image quality can be preserved across reconstructions during weeks, as long as temperature changes are small ($\pm$1~Cº) and differences in the optimal active shimming for different scans are negligible ($<$~20~µT/m).

\subsection{Standard relaxation maps}
\label{sec:goldstandardmethods}

To establish reference $T_1$ and $T_2$ values of knee tissues at 90 mT for subsequent comparison with our VFA-PETRA measurements, a preliminary experiment was performed in which a healthy knee was mapped using conventional relaxometry techniques. For the measurement of $T_2$ in the soft tissues of the knee following a standard approach (Sec.~\ref{sec:theory_relax_maps_std}), a series of RARE images were acquired with $\text{TR} \gg T_1$. The echo time (TE) was progressively varied in steps of 10\,ms, from $\text{TE}=20$\,ms up to 100\,ms, at which point the signal of all tissues became strongly attenuated. For the measurement of $T_1$, fourteen STIR images were acquired with $\text{TR} \gg T_1$, varying the inversion time (TI) from $\text{TI}=0$ up to $\text{TI}=600$\,ms, where the signals of all tissues were fully recovered from inversion. For the $T_2$ map, the volunteer was a 25-year-old male, whereas for the $T_1$ map, the volunteer was a 32-year-old male. 

In all cases, the filling of $k$-space followed a center-out trajectory with 70\,\% partial Fourier sampling in the slice direction. The $k$-spaces were FFT-reconstructed, and the noise background was removed using a common mask applied to all magnitude images. For each pixel within the mask, a non-linear least-squares fitting was performed in MATLAB using the \texttt{lsqcurvefit} function. The initial $T_2$ seeds were dynamically determined as the echo time ($TE$) at which the signal intensity first dropped below half of its initial amplitude. For $T_1$ mapping, the seeds were set to the $TI$ corresponding to the signal null point in the recovery curve. For the $T_2$ map, the magnitude data $|s(\text{TE})|$ was fitted to

\begin{equation}
	s(\rho, T_2) = \rho \, e^{-\text{TE}/T_{2}},
	\label{eq:T2fit}
\end{equation}
whereas the $T_1$ map was obtained by fitting the magnitude data $|s(\text{TI})|$ to:
\begin{equation}
	s(\rho, T_1) = \left| \rho \left( 1 - 2 e^{-\text{TI}/T_{1}} + e^{-\text{TR}/T_{1}} \right) \right|.
	\label{eq:T1fit}
\end{equation}

\subsection{VFA-PETRA relaxation maps}
\label{sec:petramethod}

Because the $T_1$ mapping strategy described in Sec.~\ref{sec:theory_relax_maps} may become inaccurate when applied to anatomical regions protruding from the RF coil bore (such as the knee or elbow), we first evaluated the method using the same homogeneous phantom employed for field mapping, which fully occupied the imaging volume (Fig.~\ref{fig:fig_scanner}b) and whose $T_1$ ground truth had previously been determined by conventional inversion recovery measurements, yielding $T_1$=12.5~ms.

This validation aimed to assess how the SNR of the resulting $T_1$ maps depends on the nominal flip angles of the PETRA pair used to compute $\lambda(\vec{r})$ in Eq.~(\ref{eq:lambdaVFAratio}), and how this relationship affects $T_1$ estimation accuracy and precision. Following phantom validation, the VFA-PETRA method was applied for \emph{in-vivo} $T_1$ mapping of the knee, simultaneously covering soft and hard tissues in a healthy volunteer. The scanned knee belonged to the same volunteer from whom the $T_1$ map was obtained using conventional methods.

The RF pulses were systematically corrected by means of pre/counter-emphasis, and the calibration of each $\alpha_\text{nom}$ was defined from Rabi flops obtained with increasing RF amplitude and fixed pulse duration, ensuring identical excitation bandwidths for both acquisitions.

For the \emph{in-vivo} case, two PETRA datasets were acquired with a fixed $\text{TR}=50$\,ms and variable nominal flip angles $\alpha_\text{1,nom}/\alpha_\text{2,nom}=10/40$° over $\text{T}_\text{seq}=6$\,min each. This configuration produced a 3D $T_1$ map with a resolution of $\Delta r = 2 \times 2 \times 4$\,mm$^3$ in a total acquisition time of 12\,min.

After the pair of PETRA acquisitions, the volunteer’s knee was removed from the RF coil and replaced with the homogeneous phantom used for field mapping (Fig.~\ref{fig:fig_scanner}b). After this exchange, we readjusted the impedance matching. Subsequently, the three SPRITE sequences described in Sec.~\ref{sec:theory_maps} were executed using the same active shimming settings as those applied during the preceding \emph{in-vivo} PETRA scans.

\subsection{Reconstruction and post-processing}
\label{sec:reconstruction}

Under ideal conditions, a radial-to-Cartesian interpolation (with optional density compensation) followed by a model-agnostic Fourier inversion can be an efficient strategy for non-Cartesian $k$-space sampling. However, in non-ideal regimes a more accurate approach is to construct a linear forward model describing the system response to the applied pulse sequence, incorporating prior knowledge of the relevant magnetic fields. The resulting model defines a cost function that is minimized (optionally with regularization) to solve the linear system of $i$ equations:
\begin{equation}\label{eq:EM}
	s_i(t) = \text{EM}_{i,j}(\vec{r},t) \, \rho_j(\vec{r}),
\end{equation}
where $s_i$ is the vectorized $k$-space signal with a length equal to the number of recorded samples during the sequence ($i$), and $\rho_j$ is the vectorized spin-density matrix with a size equal to the number of voxels in the reconstructed image ($j$). The matrix $\text{EM}_{i,j}$ represents a signal \emph{encoding matrix}. 

Once the $\Delta B_0(\vec{r})$ and $\Delta B_1(\vec{r})$ field maps have been estimated from the SPDS protocol, the corresponding field matrices (defined over the full imaging FoV) are cropped to the partial imaging region and interpolated to match the target image resolution. When reconstructing $k$-space data acquired with a PETRA sequence at a given nominal flip angle $\alpha_\text{nom}$, the voxelwise vectors $\Delta B_0(\vec{r}_j)$ and $\eta(\vec{r}_j)$ are incorporated into the non-ideal terms of the EM for each $i$-sample of $k$-space as
\begin{equation}\label{eq:encodingmatrix}
	\text{EM}_{i,j} = \epsilon(\vec{r}_j) \cdot 
	\exp\!\left\lbrace 
	-\I \left[ \vec{k}_i \cdot \vec{r}_j + \gamma \, \Delta B_0(\vec{r}_j) \, t(\vec{k}_i) \right]
	\right\rbrace,
\end{equation}
where
\begin{equation}\label{eq:gfactor}
	\epsilon(\vec{r}_j) = 
	\frac{\sin\bigl(\alpha_\text{nom} \, \eta(\vec{r}_j)\bigr)}{\sin(\alpha_\text{nom})} \cdot \eta(\vec{r}_j).
\end{equation}

On the right-hand side of Eq.~(\ref{eq:gfactor}), the left fraction represents the normalized FA experienced by each voxel during transmission, while $\eta$ accounts for the receive sensitivity of the coil at each voxel. In our PETRA acquisitions, $t(\vec{k})=T_\text{d}$ for the central pointwise region, and $t(\vec{k})=|\vec{k}|/(\gamma |g_\text{enc}|)$ for the radial portion. In RARE acquisitions, $t(\vec{k})=|\vec{k}_\text{rd}|/(\gamma |g_\text{rd}|)$.

Finally, Eq.~(\ref{eq:EM}) is inverted using the Kaczmarz algorithm (ART) \cite{kaczmarz}, based on the recursive expression
\begin{equation}\label{eq:ART}
	\rho_n =
	\rho_{n-1}
	+ \lambda \,
	\frac{
		s_i - \text{EM}_i \cdot \rho_{n-1}
	}{
		\left\| \text{EM}_i \right\|
	} \text{EM}_i^*,
\end{equation}
where $\lambda$ is a regularization parameter, $s_i$ is the $i^\text{th}$ element of the vectorized $k$-space data, $\text{EM}_i$ is the $i^\text{th}$ row of the encoding matrix, and $\rho_0$ is typically initialized to zero. The estimated solution $\rho_n$ is updated $n_\text{t} \times n_\text{it}$ times via Eq.~(\ref{eq:ART}), where $n_\text{it}$ denotes the number of ART iterations.

Except for the images shown in Fig.~\ref{fig:mapasrodillaclasicos}, which were simply FFT-inverted, all PETRA and RARE images presented here were reconstructed using ART with $B_0$ and $B_1$ PK. This decision is supported by previous studies comparing ART with other reconstruction methods for PETRA datasets acquired under different degrees of $B_0$ field homogeneity, where ART demonstrated higher SNR and greater robustness against non-stationary noise distributions (see Fig.~6 in Ref.~\cite{Algarin2020} and Fig.~S1 and S2), while maintaining superior performance under severe $B_0$ inhomogeneities without requiring additional density compensation terms (see Figs.~7 and 9 in Ref.~\cite{Borreguero2024}). In all cases, we used $\lambda = n_\text{it} = 1$. Except for the images in Figs.~\ref{fig:optimizacionpetra} and \ref{fig:fieldsmapping}c–d, all reconstructions were denoised using a Gaussian BM4D filter \cite{Maggioni2013} with $\sigma$=0. All images presented were reconstructed using a workstation equipped with an NVIDIA GeForce RTX 4070 Super GPU. In the most computationally demanding case, corresponding to the images shown in Fig.~\ref{fig:kneeisotropica} with 447,200 $k$-space samples, ART required 260~s when no PK is included ($\epsilon(\vec{r}_j)=1$ and $\Delta B_0(\vec{r}_j)=0$ in Eq.~(\ref{eq:encodingmatrix})) and 280~s when PK is considered ($\epsilon(\vec{r}_j) \neq 1$ and $\Delta B_0(\vec{r}_j) \neq 0$ in Eq.~(\ref{eq:encodingmatrix})).

\subsection{Tissue $T_1$ estimations}
\label{sec:dataanalysis}

Once the 3D $T_1$ and $T_2$ maps are obtained using either VFA-PETRA or standard methods, a systematic procedure is followed to extract the mean relaxation times for each tissue. First, a representative 2D slice of the 3D map is selected, corresponding to the region where the largest and most homogeneous portion of the tissue of interest is visible and free of artifacts. This slice is then segmented to generate a mask covering the homogeneous region of that tissue. Tissue segmentation was implemented using a semi-automated region-growing algorithm based on a 4-connected neighborhood connectivity. Starting from manually selected seed points, the regions were iteratively expanded by including adjacent voxels whose intensities remained within a predefined tolerance threshold relative to the initial seed average. However, these segmentations were not confirmed by certified radiologists, so the estimated $T_1$ values for hard tissues in Fig.\ref{fig:mapasrodillapetra}d should be interpreted with caution and considered as a preliminary approximation rather than definitive quantitative benchmarks. $T_1$ relaxation times were estimated by fitting the masked voxel-wise $T_1$ distributions inside the map to a Gaussian probability density function, defined by:

\begin{equation}\label{eq:gaussiandistribution}
f(T_1 \mid T_1^\text{mean}, \epsilon_{T_1}) = \frac{1}{\epsilon_{T_1} \sqrt{2\pi}} \exp\left( -\frac{(T_1 - T_1^\text{mean})^2}{2\epsilon_{T_1}^2} \right),
\end{equation}

using a maximum likelihood estimation (MLE) approach to extract the mean and standard deviation $(T_1^\text{mean} \pm \epsilon_{T_1})$.


\section{Results}
\label{sec:results}

\subsection{System calibrations and PETRA optimization}
\label{sec:resultsPETRAcalibrations}

The results presented below correspond to the calibration and optimization procedures described in Section~\ref{sec:met_calibration}.

\begin{figure*}
	\includegraphics[width=1.0\textwidth]{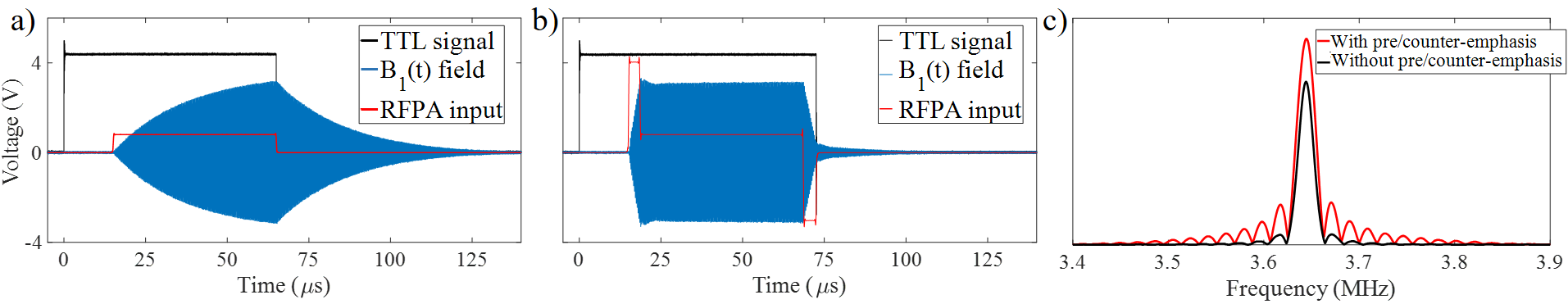}
	\caption{Temporal trace of the $B_1(t)$ field generated by the RF coil after triggering the RFPA with a pulse (a) without and (b) with pre- and counter-emphasis. (c) FFTs of the $B_1(t)$ field envelopes detected by the pickup coil in both cases.}
	\label{fig:preemphasis}
\end{figure*}

Figure~\ref{fig:preemphasis} shows the temporal evolution of the $B_1(t)$ field in the absence (a) and presence (b) of pre/counter-emphasis. Figure~\ref{fig:preemphasis}c displays the corresponding frequency spectra of the blue traces in panels a and b.

\begin{figure*}
	\includegraphics[width=1.0\textwidth]{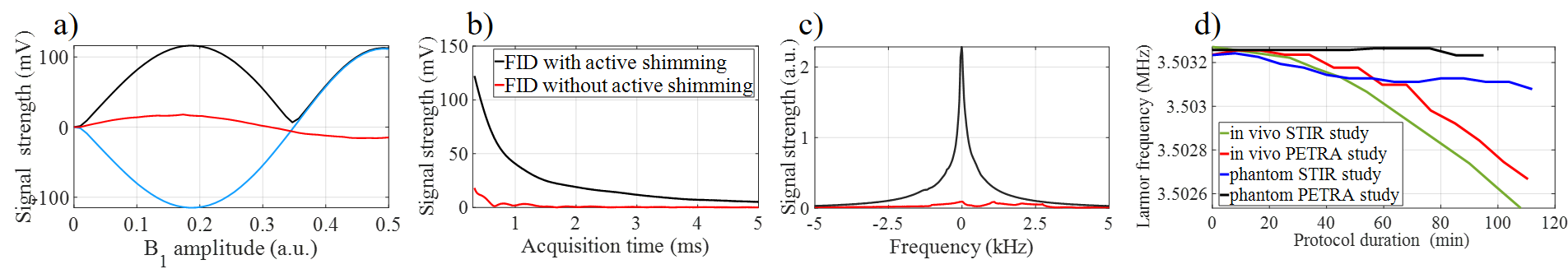}
	\caption{Calibration of the NMR signal with the RF coil loaded by a knee for PETRA optimization in Halbach systems. (a) Rabi flop acquired from the repeated application ($\text{TR}=1$\,s) of square RF pulses with pre-emphasis ($t_\text{RF}=25$\,µs) and linearly increasing amplitude for flip-angle calibration. (b) FID before and after active shimming optimization. (c) Frequency spectra of the FIDs in (b), showing line narrowing to 88\,ppm ($\text{FWHM}=320$\,Hz at 3.64\,MHz) after shimming. (d) Larmor frequency drift during PETRA and STIR studies on \emph{in-vivo} subjects and the corresponding variation observed in phantom measurements.}
	\label{fig:calibraciones}
\end{figure*}

Figure~\ref{fig:calibraciones} summarizes the calibration measurements performed prior to each experiment once the RF coil was loaded with the subject’s knee. Panel a shows a Rabi flop acquired by repeatedly applying RF pulses ($\text{TR}=1$\,s) with pre/counter-emphasis and linearly increasing amplitude. The first point of each recorded FID, acquired $T_\text{dead}=200$\,µs after switching off the RF pulse, is plotted. Panels b and c show the FIDs and their respective FFTs obtained without and with active shimming, after applying a 90° RF pulse ($t_\text{RF}=25$\,µs) with amplitude calibrated from panel a. The optimized shimming reduced the linewidth to 88\,ppm ($\text{FWHM}=320$\,Hz at 3.64\,MHz). Panel d shows the Larmor frequency drift measured during long PETRA and STIR acquisitions \emph{in-vivo} (corresponding to Figs.~\ref{fig:mapasrodillaclasicos}b and \ref{fig:mapasrodillapetra}a) and during analogous phantom experiments.

\begin{figure}
	\includegraphics[width=0.5\textwidth]{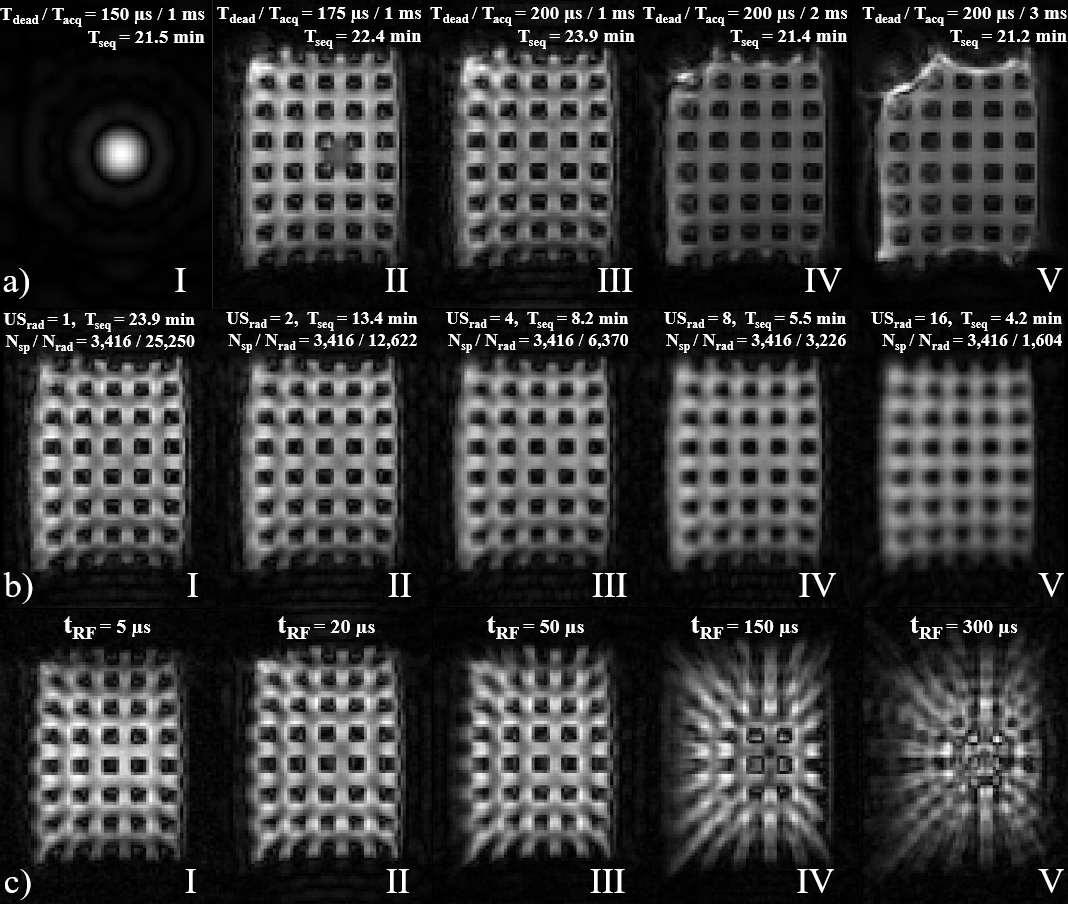}
	\caption{Evaluation of reconstruction quality of a reticulated phantom encoded with PETRA sequences: (a) effect of the ratio between Tx/Rx switching time and acquisition window ($T_\text{dead}/T_\text{acq}$), (b) effect of radial $k$-space undersampling ($\text{US}_\text{rad}$), and (c) dependence on RF pulse length ($t_\text{RF}$).}
	\label{fig:optimizacionpetra}
\end{figure}

Figure~\ref{fig:optimizacionpetra}a illustrates the influence of the Tx/Rx switching time ($T_\text{dead}$) and acquisition window ($T_\text{acq}$) on PETRA reconstructions. From left to right, the ratio $T_\text{dead}/T_\text{acq}$ varies around the optimal value found for this phantom, set to 200\,µs/1\,ms. Figure~\ref{fig:optimizacionpetra}b shows the impact of radial $k$-space undersampling ($\text{US}_\text{rad}$). As $\text{US}_\text{rad}$ increases from $\times1$ to $\times16$, the sequence duration shortens at the expense of nominal resolution. Figure~\ref{fig:optimizacionpetra}c shows the artifacts caused by RF pulse length ($t_\text{RF}$) in PETRA images acquired with $|g_\text{enc}|=6$\,mT/m and a bandwidth 38\,kHz. From left to right, $t_\text{RF}$ increases as 5, 20, 50, 150, and 300\,µs, corresponding to spectral bandwidths of 240, 60, 24, 8, and 4\,kHz, respectively.

\subsection{Field maps and model-based reconstruction}
\label{sec:resultsfieldsmapping}

The results presented here correspond to the procedures described in Section~\ref{sec:fieldmapping} (field mapping) and Section~\ref{sec:reconstruction} (reconstruction).

\begin{figure*}
	\includegraphics[width=1.0\textwidth]{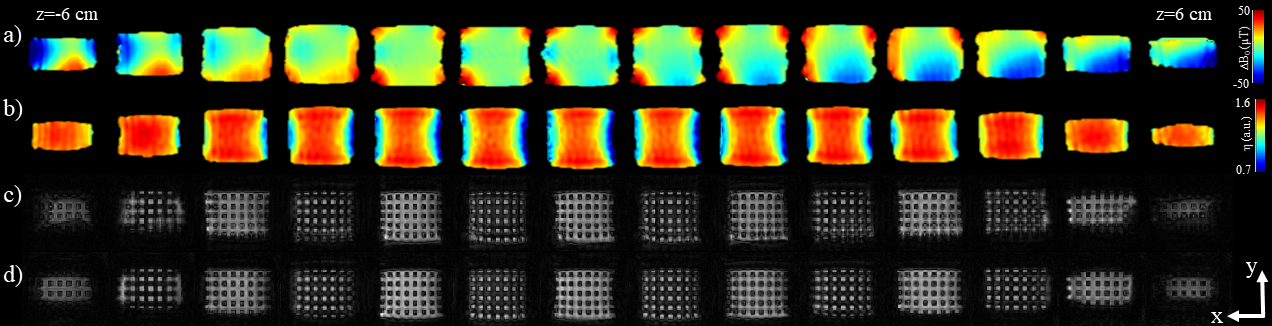}
	\caption{Evaluation of $B_0(\vec{r})$ and $B_1(\vec{r})$ mapping protocols and comparison between agnostic and model-based reconstruction methods. 
		(a) Different slices along the $z$-direction of the $\Delta B_0(\vec{r})$ map. 
		(b) Corresponding slices showing the $B_1$ efficiency map $\eta(\vec{r})$. 
		(c) Same slices reconstructed with ART without PK of the fields (Eq.~(\ref{eq:EM}) with $\Delta B_0(\vec{r}_j)=0$ and $\epsilon(\vec{r}_j)=1$) from a grid phantom encoded with PETRA. 
		(d) Same slices reconstructed with ART including PK of both $B_0(\vec{r})$ and $B_1(\vec{r})$, using the same PETRA sequence as in (c).}
	\label{fig:fieldsmapping}
\end{figure*}

Figures~\ref{fig:fieldsmapping}a and b show, respectively, the $\Delta B_0(\vec{r})$ and $\eta(\vec{r})$ maps obtained using the SPDS protocol for several $z$-slices across the FoV. 
Figures~\ref{fig:fieldsmapping}c and d compare agnostic and model-based reconstructions of the same PETRA $k$-space data acquired from a gridded phantom filled with a 1\,\% CuSO$_4$ solution.

\subsection{Standard relaxation maps}
\label{sec:std_maps}

The results presented below correspond to the standard procedures described in Section~\ref{sec:goldstandardmethods} for $T_1$ and $T_2$ mapping.

\begin{figure}
	\includegraphics[width=0.5\textwidth]{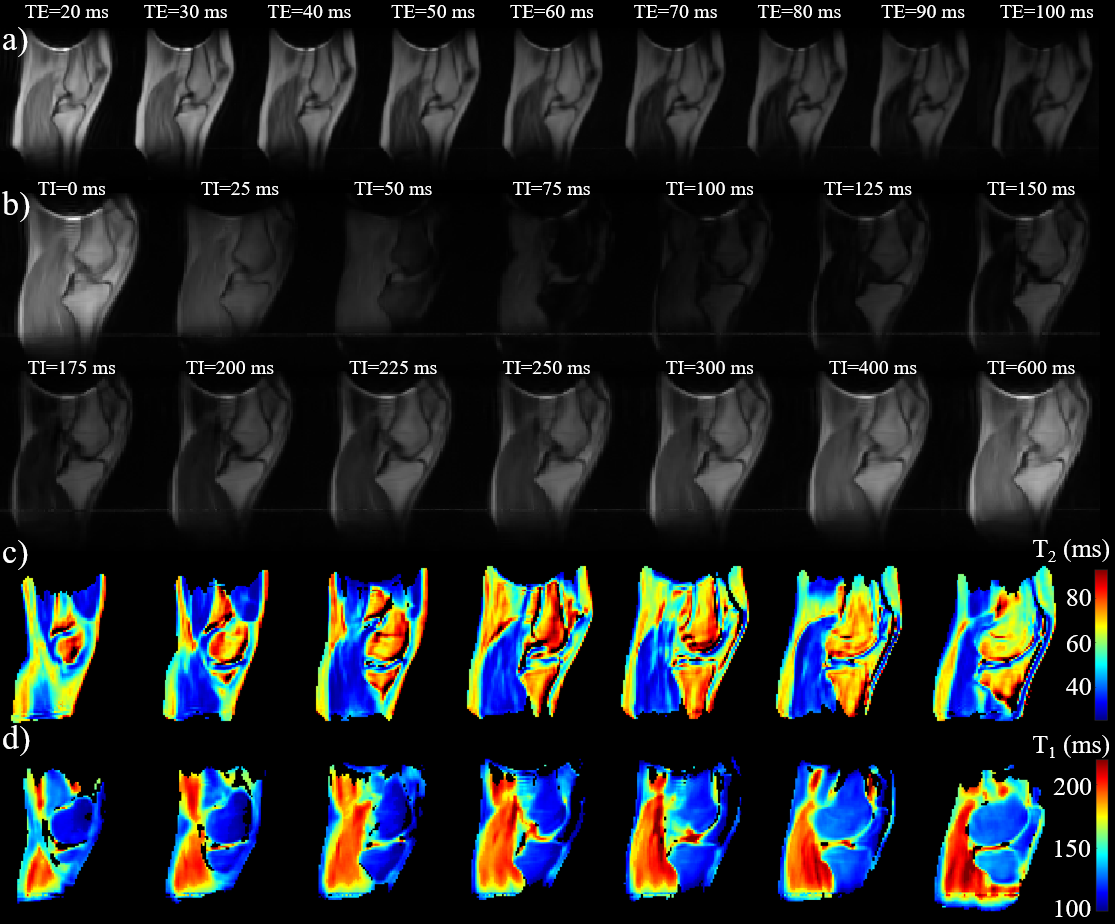}
	\caption{Quantitative $T_1$ and $T_2$ maps obtained using gold-standard methods. 
		(a) Single slice containing the posterior cruciate ligament from a series of nine RARE images acquired at varying echo times (TE). 
		(b) Same slice from a series of fourteen STIR images acquired at varying inversion times (TI). 
		(c) $T_2$ map computed from the data in (a). 
		(d) $T_1$ map computed from the data in (b).}
	\label{fig:mapasrodillaclasicos}
\end{figure}

Figure~\ref{fig:mapasrodillaclasicos}a shows a representative slice from a series of nine 3D RARE images of a healthy volunteer’s knee, acquired with echo times ranging from 10\,ms to 100\,ms. 
Figure~\ref{fig:mapasrodillaclasicos}b shows the same slice from a series of fourteen 3D STIR images acquired in another healthy volunteer, with inversion times ranging from 0\,ms to 600\,ms. 
Figures~\ref{fig:mapasrodillaclasicos}c and d display the corresponding $T_2$ and $T_1$ maps of the knee’s soft tissues, obtained by voxelwise fitting of the signal variation with TE and TI according to Eqs.~(\ref{eq:T2fit}) and~(\ref{eq:T1fit}), respectively.

\subsection{Phantom validation of VFA-PETRA relaxation maps}
\label{sec:resultscalibrationVFAPETRA}

The results presented below correspond to procedures described in Section~\ref{sec:petramethod}.

\begin{figure*}
	\includegraphics[width=1.0\textwidth]{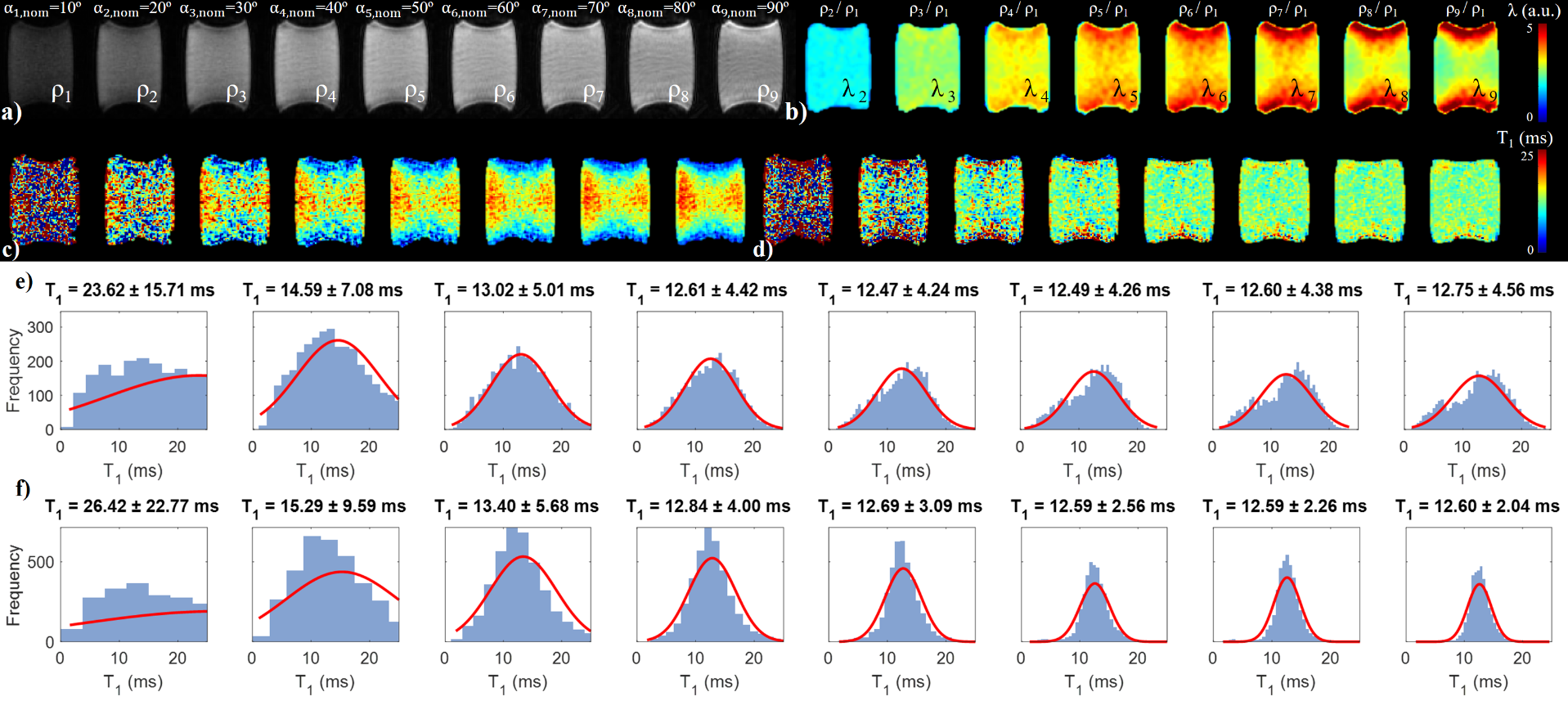}
	\caption{Validation of VFA-PETRA $T_1$ mapping on a homogeneous phantom with nominal $T_1$=12.5~ms determined by simple inversion recovery. 
		(a) ART reconstructions with $B_0$ PK of nine PETRA sequences acquired with different nominal FAs. 
		(b) Ratio maps $\lambda_{i}(\vec{r})$ computed using $\rho_1$ as the baseline for the lowest-FA image. 
		(c) $T_1$ maps obtained by fitting Eq.~(\ref{eq:lambdaVFAratio}) without PK of $B_0$ or $B_1$. 
		(d) $T_1$ maps obtained from the same data when including $B_0$ and $B_1$ PK in the fitting model.
		(e) Histograms of $T_1$ distributions for each map in (c).
		(f) Histograms of $T_1$ distributions for each map in (d).}
	\label{fig:calibracionVAFPETRA}
\end{figure*}

Figure~\ref{fig:calibracionVAFPETRA} shows the phantom-based validation of the VFA-PETRA approach for $T_1$ mapping, evaluating its reliability as a function of: (i)~the nominal flip angles used for the PETRA sequence pair; and (ii)~the inclusion or omission of $B_0$ and $B_1$ PK in the fitting model. 

Figure~\ref{fig:calibracionVAFPETRA}a displays the central $z$-slice from a set of nine PETRA images ($\rho_i$, $i=1\dots9$) acquired with nominal flip angles varying from $\alpha_\text{1,nom}=10$° to $\alpha_\text{9,nom}=90$° in 10° increments. All images were reconstructed with ART using only $B_0$ PK in the encoding matrix. 

Figure~\ref{fig:calibracionVAFPETRA}b shows the corresponding ratio maps $\lambda_{i}(\vec{r}) = \text{abs}(\rho_{i}) / \text{abs}(\rho_{1})$ ($i=2\dots9$) computed for the same slice. 

Figure~\ref{fig:calibracionVAFPETRA}c presents the $T_1$ maps obtained from each $\lambda_{i}$ by fitting Eq.~(\ref{eq:lambdaVFAratio}) under the assumption of negligible field inhomogeneities ($\Delta B_0=0$, $\eta=1$). 
Figure~\ref{fig:calibracionVAFPETRA}d shows the corresponding $T_1$ histograms derived from these maps. 

Figure~\ref{fig:calibracionVAFPETRA}e displays the $T_1$ maps obtained from the same ratios when including the $\Delta B_0$ and $\eta$ maps estimated from SPDS as PK in the fitting model. 
Finally, Figure~\ref{fig:calibracionVAFPETRA}f shows the $T_1$ histograms corresponding to these corrected maps.

\subsection{In-vivo imaging of hard tissues}
\label{sec:resultspetraenhalbach}

The results presented here correspond to the PETRA sequences detailed in Table~\ref{tab:PETRAimageParameters}, after applying the optimizations and calibrations described in the previous subsections.

\begin{figure*}
	\includegraphics[width=1.0\textwidth]{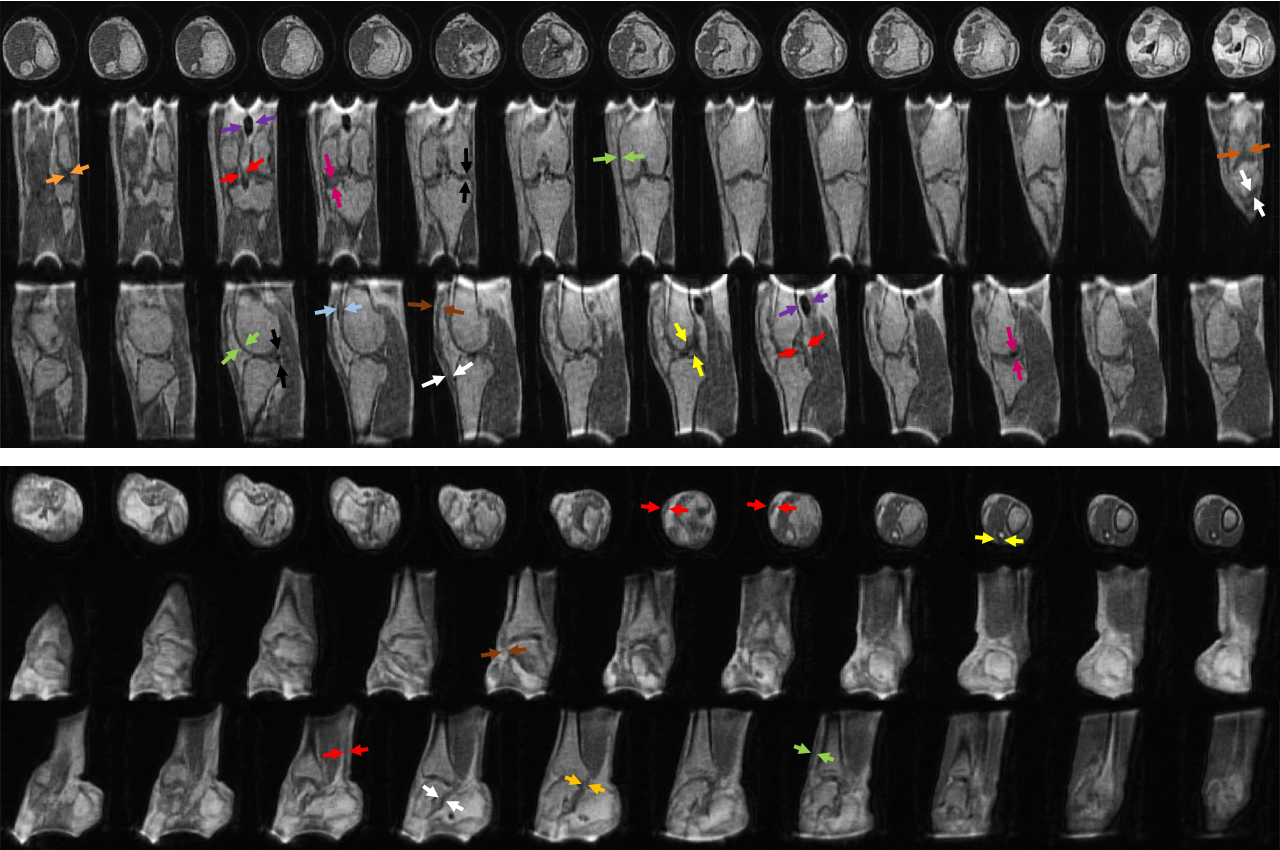}
	\caption{Axial, coronal, and sagittal slices reconstructed from PETRA acquisitions of the knee (top) and ankle (bottom) of a volunteer. In the knee, identifiable structures include the anterior (yellow) and posterior (red) cruciate ligaments, patellar cartilage (blue), cortical bone (green), quadriceps tendon (brown), patellar ligament (white), fibular collateral ligament (orange), and the medial (pink) and lateral (black) menisci. A Baker’s cyst (purple) appears dark. In the ankle: Achilles tendon (red), peroneus longus tendon (yellow), tibialis anterior tendon (green), posterior tibiofibular ligament (orange), tibialis posterior tendon (brown), and flexor hallucis tendon (white).}
	\label{fig:kneeisotropica}
\end{figure*}

Figure~\ref{fig:kneeisotropica} shows reconstructions of $k$-space data acquired for the knee and ankle of a healthy volunteer. 
For consistent visualization across all three anatomical planes, an isotropic spatial resolution of 2\,mm was selected, despite the associated penalty in SNR and scan time.

\begin{figure}
	\includegraphics[width=0.5\textwidth]{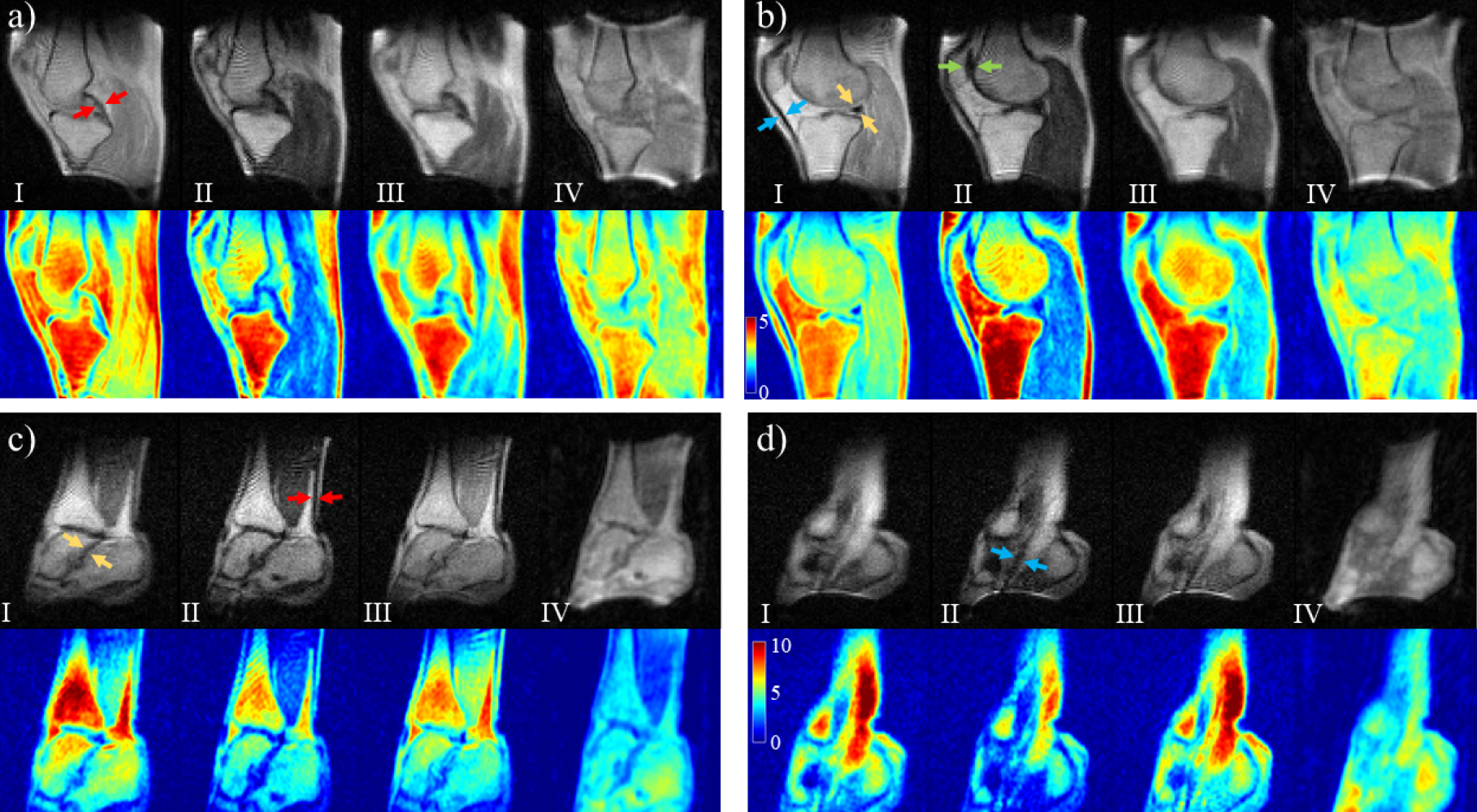}
	\caption{Qualitative comparison of the same slice of a healthy subject’s knee (top) and ankle (bottom) imaged with four different sequences, together with their corresponding SNR maps. From left to right: RARE-PD\textsubscript{w}, RARE-$T_1$\textsubscript{w}, RARE-$T_2$\textsubscript{w}, and PETRA. In the knee: anterior cruciate ligament (red), patellar tendon (green), patellar cartilage (red), and lateral meniscus (yellow). In the ankle: Achilles tendon (red), posterior talofibular ligament (yellow), and peroneus longus tendon (blue). SNR maps were generated by computing the ratio between the local mean intensity (sliding window of $3\times3$ pixels) and the background noise standard deviation, normalized by $\sqrt{T_{scan}}$.}
	\label{fig:rarevspetra}
\end{figure}

Figure~\ref{fig:rarevspetra} compares the visualization capabilities of PETRA with those of standard clinical MRI sequences (PD-, $T_1$-, and $T_2$-weighted RARE) for both the knee and ankle. 
Below each anatomical slice, the corresponding SNR map is shown, normalized per unit of scan time and displayed using a common color scale.

\begin{figure}
	\includegraphics[width=0.5\textwidth]{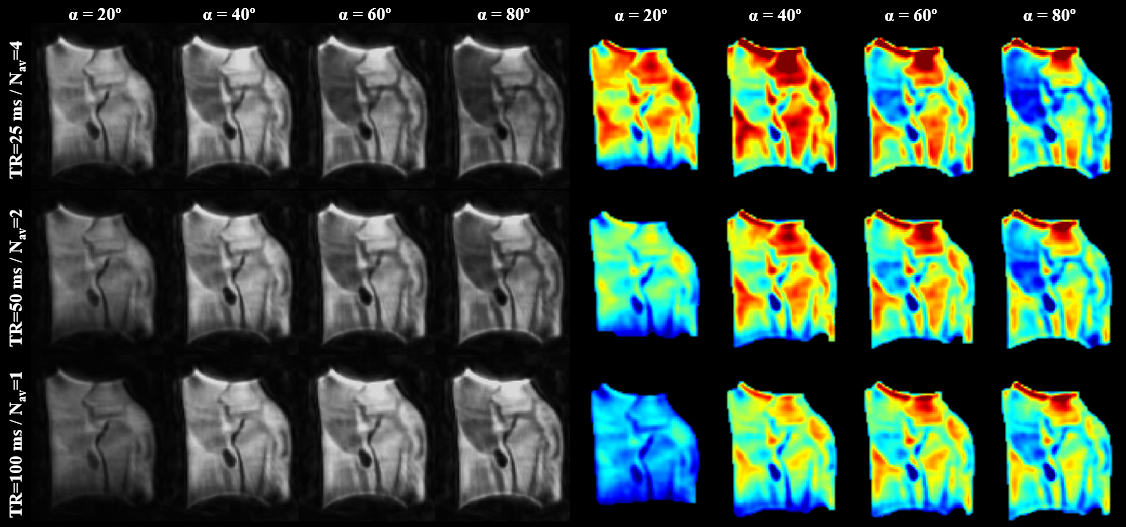}
	\caption{Reconstructions and SNR maps of the same $z$-slice of a knee, acquired with equivalent PETRA sequences at different flip angles (FA) and repetition times (TR).}
	\label{fig:FAandTRsweep}
\end{figure}

Figure~\ref{fig:FAandTRsweep} shows the same $z$-slice of a healthy volunteer’s knee, encoded with equivalent PETRA sequences differing only in FA and TR. The number of averages was fixed for each acquisition to maintain a constant total scan duration (8.4\,min).

\begin{figure}
	\includegraphics[width=0.5\textwidth]{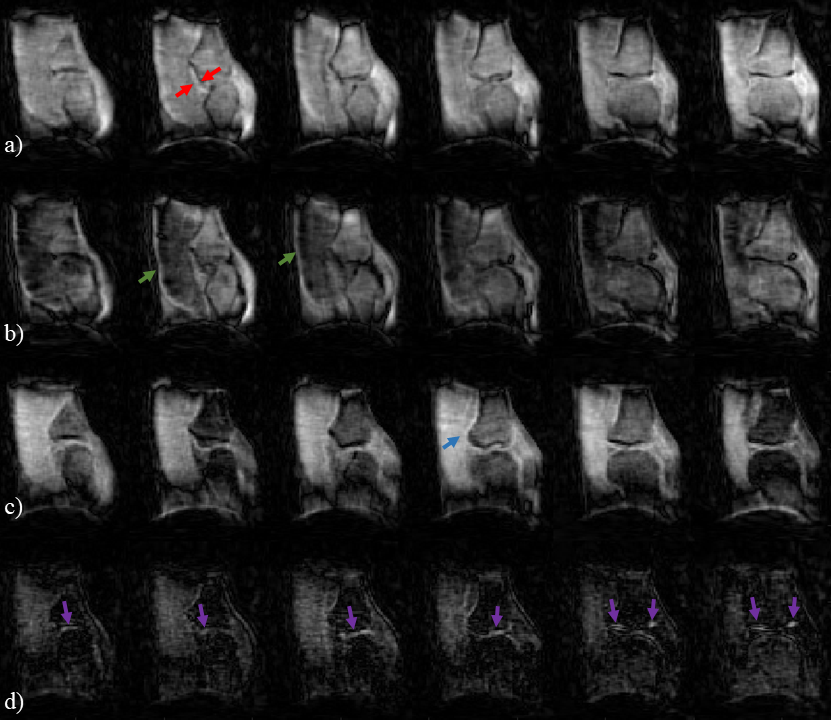}
	\caption{Slices from four 3D datasets generated by subtraction of PETRA images acquired at different flip angles. 
		(a) Hyperintense visualization of the cruciate ligament above muscle (red). 
		(b) Hyperintense visualization of fat relative to muscle (green). 
		(c) Fat suppression and enhanced muscle contrast (blue). 
		(d) Hyperintense visualization of cartilage and meniscus (purple).}
	\label{fig:contrast}
\end{figure}

Figure~\ref{fig:contrast} shows slices from four 3D datasets obtained by subtracting PETRA images acquired with fixed $\text{TR}=50$\,ms and variable flip angles ranging between 10° and 150°. 
Figure~\ref{fig:contrast}a corresponds to $\alpha_2 - \alpha_1 = 120^\circ - 50^\circ$, highlighting the cruciate ligament (red arrow). 
Figure~\ref{fig:contrast}b corresponds to $\alpha_2 - \alpha_1 = 150^\circ - 90^\circ$, enhancing fat relative to muscle (green arrow). 
Figure~\ref{fig:contrast}c corresponds to $\alpha_2 - \alpha_1 = 70^\circ - 20^\circ$, showing complete fat suppression and muscle enhancement (blue arrow). 
Finally, Figure~\ref{fig:contrast}d corresponds to $\alpha_2 - \alpha_1 = 30^\circ - 10^\circ$, emphasizing cartilage and meniscus (purple arrow).

\subsection{In-vivo relaxation maps of hard tissues}
\label{sec:resultsT1mapping}

The results presented below correspond to methods described in Section~\ref{sec:petramethod} and validated as shown in Section~\ref{sec:resultscalibrationVFAPETRA}.

\begin{figure*}
	\includegraphics[width=1\textwidth]{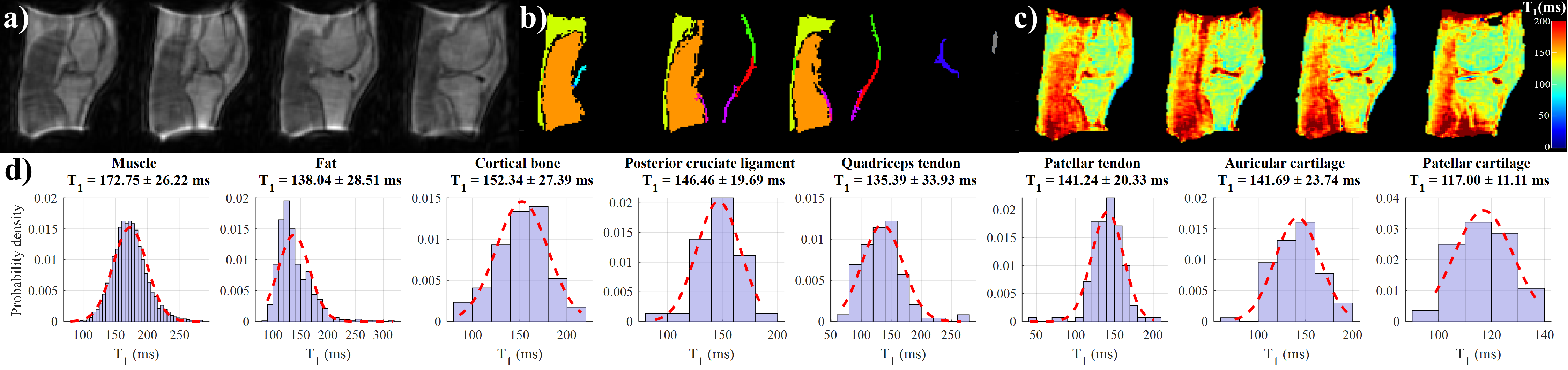}
	\caption{\emph{in-vivo} $T_1$ mapping of a healthy volunteer’s knee using VFA-PETRA. 
		(a) Selected slices for $T_1$ analysis. 
		(b) Segmentation of regions of interest (RoIs) for each tissue type. 
		(c) Resulting $T_1$ maps for the same slices. 
		(d) $T_1$ distribution within each segmented RoI and corresponding Gaussian fits for mean $T_1$ estimation.}
	\label{fig:mapasrodillapetra}
\end{figure*}

Figure~\ref{fig:mapasrodillapetra} shows the results of 3D $T_1$ mapping performed with VFA-PETRA on the knee of a healthy volunteer. 
Figure~\ref{fig:mapasrodillapetra}a displays four representative slices from a 3D PETRA acquisition encompassing the main tissues of interest. 
This image was generated using PETRA parameters equivalent to those employed for the pair of images used to compute $\lambda(\vec{r})$ in Eq.~(\ref{eq:lambdaVFAratio}), except for the flip angle, which was set to 90° to enhance tissue contrast and facilitate automatic segmentation. 

Figure~\ref{fig:mapasrodillapetra}b shows the segmentation of short-$T_2$ tissues in the same slices, from which the mean $T_1$ value for each region was subsequently determined. 
The segmented structures include the posterior cruciate ligament (sky blue), quadriceps tendon (green), patellar tendon (red), cortical bone (purple), auricular cartilage (dark blue), and patellar cartilage (gray). 
For comparison with standard methods, two predominant soft tissues (muscle and fat, shown in orange and yellow, respectively) were also included. 

Figure~\ref{fig:mapasrodillapetra}c presents the corresponding $T_1$ maps for the same slices, obtained from $\lambda(\vec{r})$ computed using a pair of PETRA images with nominal flip angles $\alpha_\text{1,nom}/\alpha_\text{2,nom}=10^\circ/40^\circ$, followed by voxelwise fitting to Eq.~(\ref{eq:facalibration}). 
Finally, Figure~\ref{fig:mapasrodillapetra}d shows the $T_1$ distributions within each RoI together with their Gaussian fits, from which the mean $T_1$ values for each tissue were estimated.


\section{Discussion}
\label{sec:discusion}

\subsection{System calibrations and PETRA optimization}
\label{sec:discusioncalibratonspetra}

Figure~\ref{fig:preemphasis}a shows the distortion of the RF pulse relative to the expected rectangular shape. In this case, the charging and discharging of energy in the resonator extend up to $\tau \approx 22$\,µs due to the low operating frequency ($\omega_\text{L}=3.64$\,MHz) and the high quality factor ($Q\approx250$) of the RF coil. However, Figure~\ref{fig:preemphasis}b demonstrates that this distortion can be effectively compensated by incorporating pre-emphasis, while the counter-emphasis reduces spurious energy stored in the resonator. Figure~\ref{fig:preemphasis}c further shows that the application of pre/counter-emphasis enables a more uniform excitation of a 24\,kHz spectral bandwidth.

In Figure~\ref{fig:calibraciones}a, the uncertainty in flip-angle calibration derived from a Rabi flop is evident, originating from $B_1$ inhomogeneities. This manifests as a non-zero signal at $B_1=0.35$\,a.u., corresponding to the $\pi$-pulse condition. Figures~\ref{fig:calibraciones}b and c show that active shimming extends the raw FID up to $T_2^* \approx 1$\,ms, while narrowing the spectral width to 320\,Hz (88\,ppm). 

The Larmor frequency drift observed across different studies (Fig.~\ref{fig:calibraciones}d) supports two conclusions: (i) the PETRA sequences used in this study do not induce greater heating than the RARE/STIR sequences; and (ii) the drift in Larmor frequency is primarily driven by heat transfer from the subject to the magnet, together with the sensitivity of the permanent magnets to ambient conditions. During \emph{in-vivo} knee experiments, a maximum shift of 650\,Hz (corresponding to a $B_0$ fluctuation of 15~µT) was recorded after 110\,min of consecutive acquisitions, all performed under identical gradient conditions. This corresponds to a a Larmor drift of approximately 80 ppm/h, which is an order of magnitude smaller than the intrinsic $B_0$ inhomogeneity of the system. Consequently, the impact of temperature-induced $B_0$ fluctuations on the measured $T_1$ relaxation times is considered negligible within our experimental setup. These thermal fluctuations are also not expected to significantly distort the spatial profiles of the $B_0$ and $B_1$ field maps (Fig.~\ref{fig:fieldsmapping}a,b), thereby justifying their reuse over several days under controlled ambient conditions ($\approx22\pm1$\,$^\circ$C). Furthermore, substantial variations in $B_0(r)$ and $B_1(r)$ would have compromised the consistency check shown in Fig.~\ref{fig:calibracionVAFPETRA}d, where the predicted mean value of $T_1(r)=12.5$\,ms agrees with the inversion recovery measurement.

Figure~\ref{fig:optimizacionpetra}a reveals that excessively short $T_\text{dead}$ values produce artifacts in the form of a central lobe resembling a sinc function, caused by residual ring-down energy. The required dead time depends primarily on the signal amplitude, which varies with the size of the anatomical region, as well as with image resolution, TR, and FA. When $T_\text{acq} > T_2^* \approx 1$\,ms, blurring, anisotropic resolution loss, geometric distortions, and hyperintense banding appear as a consequence of incorrect spin mapping induced by $B_0(\vec{r})$ inhomogeneities. Unlike $T_\text{dead}$, this limit is intrinsic to the system, since $T_2^*$ is mainly determined by the distribution of Larmor frequencies within the imaging volume. For these reasons, while $T_\text{acq}$ was fixed at 1.25\,ms for \emph{in-vivo} imaging, $T_\text{dead}$ was adjusted from 175\,µs (knee) to 300\,µs (ankle), reflecting the lower signal amplitude in the latter.

Figure~\ref{fig:optimizacionpetra}b illustrates that reducing the number of radial readouts in $k$-space leads to a loss of nominal resolution, which is less noticeable in highly structured phantoms such as the one used here. We found that undersampling up to a factor of $\times4$ remains acceptable for large anatomical structures, introducing only minor degradation in fine details. 

Finally, Figure~\ref{fig:optimizacionpetra}c shows that extending the RF pulse in ZTE introduces two distinct effects: (i) the non-uniform spectral coverage of long pulses produces a sinc-like radial modulation of brightness in the phantom, as peripheral regions are less efficiently excited; and (ii) the spectral selectivity of long pulses under a gradient field promotes adiabatic-like excitation, introducing a relative time-origin shift in $k(t)$ between isochromats. This effect results in image blurring in regions where spins are significantly detuned from the RF carrier frequency.

\subsection{Field maps and model-based reconstruction}

Figures~\ref{fig:fieldsmapping}a and b demonstrate the capability of the SPDS extension to generate simultaneous $B_0(\vec{r})$ and $B_1(\vec{r})$ maps that are free from geometric distortions. In the resulting maps, the expected $B_1$ modulation toward the edges of the solenoidal coil and the characteristic high-order field components of the Halbach magnet are markedly visible. 

Figure~\ref{fig:fieldsmapping}c shows the types of artifacts typically observed in PETRA reconstructions when field-agnostic methods are applied. These include geometric distortions, pronounced blurring, and complete signal loss in regions most affected by strong $B_0$ inhomogeneity. 

In contrast, Figure~\ref{fig:fieldsmapping}d shows a substantial reduction of these artifacts when PK of both $B_0(\vec{r})$ and $B_1(\vec{r})$ is incorporated into the EM within the ART reconstruction, demonstrating the advantage of the model-based approach.

\subsection{Phantom validation of VFA-PETRA relaxation maps}

Figure~\ref{fig:calibracionVAFPETRA} validates the VFA-PETRA approach for $T_1$ quantification in tissues with short $T_2$. As shown in Figure~\ref{fig:calibracionVAFPETRA}b, increasing the difference between the nominal flip angles of the PETRA image pair enhances the SNR of the $\lambda(\vec{r})$ map, thereby improving the robustness of the $T_1$ estimation. However, this also amplifies the effects of $B_0$ and $B_1$ fluctuations, which in turn demand accurate field mapping via SPDS.

Figure~\ref{fig:calibracionVAFPETRA}c illustrates that, when $\lambda(\vec{r})$ maps with large FA separations are used to generate $T_1$ maps without accounting for field inhomogeneities in Eq.~(\ref{eq:lambdaVFAratio}), the resulting $T_1$ map exhibits good SNR but deviates substantially from the expected homogeneous distribution of this phantom. Such bias could compromise the precision of $T_1$ estimates for tissues located away from the coil center in \emph{in-vivo} VFA-PETRA applications. Conversely, when a small FA difference is selected, the low SNR reduces the apparent $B_1(\vec{r})$-related inhomogeneity in the resulting $T_1$ map, but the estimation becomes unreliable due to insufficient signal contrast.

Finally, Figure~\ref{fig:calibracionVAFPETRA}d shows that, when PK of both $B_0$ and $B_1$ fields is incorporated into Eq.~(\ref{eq:lambdaVFAratio}) and a large $\lambda(\vec{r})$ map is used, the resulting $T_1$ distribution becomes spatially homogeneous across the entire phantom while retaining SNR. This improvement, which stems from the inclusion of $B_0$ and $B_1$ PK and the choice of sufficiently separated nominal flip angles, is reflected in the mean values and standard deviations of the Gaussian fits displayed in Figures~\ref{fig:calibracionVAFPETRA}e and f. However, distortions remain at the upper and lower phantom boundaries due to gradient nonlinearities, since the reconstructions did not incorporate PK of the simulated gradient fields \cite{Koolstra2021}. This simplification was justified because these nonlinearities primarily affect geometry rather than image intensity, which is the main factor influencing accurate $T_1$ mean estimation. Regardless, the intricate dependence of $M_\text{\emph{i},ss}$ on the estimated $B_1(r)$ and $T_1(r)$ values, together with the convergence toward a unique $T_1$ value for the homogeneous sample consistent with the inversion-recovery measurement, suggests that the proposed procedure is robust against potential sources of error, including operation within an incoherent regime.

\subsection{In-vivo imaging of hard tissues}
\label{sec:discussionpetraenhalbach}

The slices shown in Figure~\ref{fig:kneeisotropica} highlight several hard-tissue structures exhibiting non-zero signal intensity in the PETRA reconstructions. In the knee, identifiable structures include the anterior (yellow) and posterior (red) cruciate ligaments, patellar cartilage (blue), cortical bone (green), quadriceps tendon (brown), patellar ligament (white), fibular collateral ligament (orange), and the medial (pink) and lateral (black) menisci. A Baker’s cyst (purple) appears dark due to its long $T_1$ (greater than 1\,s) compared with the short repetition time used (TR\,=\,20\,ms). In the ankle slices, distinct structures can be identified, such as the Achilles tendon (red), peroneus longus tendon (yellow), tibialis anterior tendon (green), posterior tibiofibular ligament (orange), tibialis posterior tendon (brown), and flexor hallucis tendon (white).

Figure~\ref{fig:rarevspetra} compares the performance of PETRA against RARE sequences (PD-weighted, $T_1$-weighted, and $T_2$-weighted) for imaging the knee and ankle. In the knee slices, the image in Figure~\ref{fig:rarevspetra}a contains the largest extent of the anterior cruciate ligament (red), while Figure~\ref{fig:rarevspetra}b highlights the patellar tendon (green), patellar cartilage (red), and lateral meniscus (yellow). In the ankle, Figure~\ref{fig:rarevspetra}c shows the slice containing the widest portion of the Achilles tendon (red) and posterior talofibular ligament (yellow), whereas Figure~\ref{fig:rarevspetra}d displays the peroneus longus tendon (blue). 

A comparison of the SNR maps (normalized per unit of scan time and shown below each image) reveals that PETRA achieves substantial SNR in regions where RARE sequences show little or no signal. Among the RARE variants, PD-weighted imaging provides the closest performance to PETRA, since shorter TE and ETL values combined with longer TRs improve the detectability of tissues with short $T_2$, long $T_1$, and low proton density.

Figure~\ref{fig:FAandTRsweep} illustrates the dependence of tissue contrast and SNR efficiency in PETRA on flip angle and repetition time. According to the Ernst equation, the maximum SNR for fat and muscle (assuming an average $T_1 = 150$\,ms) occurs at $\alpha \approx 30$° for TR\,=\,25\,ms and at $\alpha \approx 60$° for TR\,=\,100\,ms. However, as in steady-state spin-echo sequences, the most efficient SNR performance is achieved at the shortest possible TR when the FA is adjusted close to the corresponding Ernst angle (in our case, TR\,=\,25\,ms and $\alpha = 40$°).

Figure~\ref{fig:contrast} demonstrates the effectiveness of applying VFA strategies in PETRA imaging to modulate tissue contrast in a controlled and predictable manner. By selecting appropriate pairs of flip angles, preferential contrast enhancement can be achieved between specific structures, as illustrated in each subfigure: cruciate ligament, fat, muscle, cartilage, and meniscus are selectively highlighted depending on the chosen flip-angle combination. This targeted contrast notably improves the anatomical delineation achievable with ZTE, which is otherwise largely limited to PD- or $T_1$-weighted contrast. Simultaneously, acquiring two PETRA images with calibrated flip angles enables the subsequent computation of $T_1$ relaxation maps (see Fig.~\ref{fig:mapasrodillapetra}c), thereby integrating qualitative contrast manipulation and quantitative relaxometry within a single acquisition framework. This underscores both the versatility and the potential diagnostic value of combining PETRA with VFA, in contrast to traditional inversion-recovery approaches.

\begin{figure*}
	\includegraphics[width=1.0\textwidth]{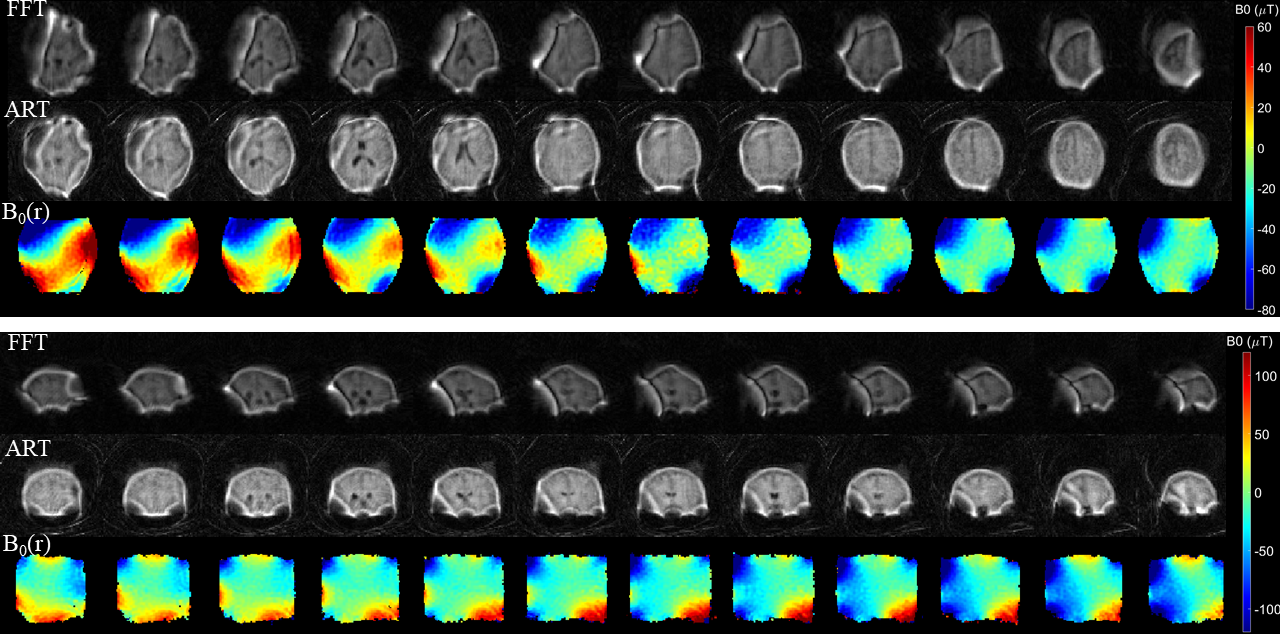}
	\caption{Axial (top) and coronal (bottom) slices of a human brain acquired with PETRA on the 90\,mT Halbach scanner. The upper rows show the FFT reconstruction, whereas the middle rows display the model-based ART reconstruction. The bottom rows present the $B_0(r)$ maps obtained using SPDS with a customized phantom (filled with a 16\% CuSO$_4$ aqueous solution) designed to occupy a volume equivalent to that of the human head. The measured spatial $B_0$ information was subsequently fitted to a sixth-order polynomial function and incorporated into the encoding matrix (Eq. 12) used for the ART reconstruction. }
	\label{fig:brainpetra}
\end{figure*}

In these \emph{in-vivo} PETRA acquisitions, several residual artifacts remain visible. The most prominent are: 1) correlated background noise associated with the Tx/Rx switching gap, despite the use of RF counter-emphasis; 2) resolution loss and blurring at tissue interfaces due to $T_\text{acq}>T_2^{*}$, despite the careful shimming performed; 3) wave-like patterns with varying orientations, likely originating from minor current instabilities in the gradient coils or limitations of the gradient power amplifier (GPA); 4) partial-volume effects obscuring smaller structures, resulting from the slice thickness selected to maintain a clinically acceptable $T_\text{scan}$; and 5) hyper-intense bands near the boundaries of the gradient linearity region, which persist despite the model-based reconstruction employed. Artifacts of the latter type become particularly exacerbated when PETRA sequences are employed for brain imaging in Halbach systems (see Fig.~\ref{fig:brainpetra}), since the patient’s shoulders prevent positioning the head at the center of the scanner. As a result, the anatomy to imaging is located not only outside the linear gradient region, but also within regions of substantially more severe $B_0$ field inhomogeneity. All these artifacts become substantially more pronounced in Fig.~\ref{fig:FAandTRsweep}, since the direct subtraction of images propagates the intrinsic artifacts of both source datasets into the resulting images. Furthermore, the circular and vertical halos visible in the axial and coronal/sagittal knee slices in Fig.~\ref{fig:kneeisotropica}, respectively, are not intrinsic to the PETRA sequence, but instead arise from the PLA holder used for the RF coil winding. In contrast, no evidence of eddy-current, chemical-shift, or patient-motion artifacts was observed.

\subsection{In-vivo relaxation maps of soft and hard tissues}
\label{sec:discussionT1map}

From the analysis of the data in Figures~\ref{fig:mapasrodillaclasicos}c and d, and according to Eqs.~(\ref{eq:T2fit}) and (\ref{eq:T1fit}), the muscle exhibits relaxation times of $T_1 = 189 \pm 16$\,ms and $T_2 = 35 \pm 5$\,ms, while fat shows $T_1 = 128 \pm 11$\,ms and $T_2 = 69 \pm 7$\,ms. These values are consistent with those reported in Ref.~\cite{OReilly2021} at 50\,mT, who measured $T_1 = 183 \pm 10$\,ms and $T_2 = 41 \pm 3$\,ms for muscle, and $T_1 = 134 \pm 7$\,ms and $T_2 = 104 \pm 10$\,ms for fat. As expected, this conventional approach yields highly reliable $T_1$ and $T_2$ estimates for soft tissues, although it remains time-consuming and impractical for routine clinical use. Nonetheless, it provides a valuable reference for assessing the accuracy of the $T_1$ values obtained using VFA-PETRA.

Figure~\ref{fig:calibracionVAFPETRA} demonstrates the capability of VFA-PETRA to generate $T_1$ maps of short-$T_2$ phantoms ($T_2 \approx 5$\,ms), which are inaccessible using standard relaxation mapping techniques. However, it is essential to include $B_1(\vec{r})$ information in the theoretical model to ensure robust estimation. As shown in Figure~\ref{fig:calibracionVAFPETRA}c, neglecting $B_1$ fluctuations results in inhomogeneous $T_1$ maps that mirror the $B_1$ efficiency distribution, leading to strong bias and poor precision in Gaussian fits. Conversely, when $B_1$ PK is incorporated (Figure~\ref{fig:calibracionVAFPETRA}d), the $T_1$ maps display homogeneous distributions, and the fitted values match the expected reference with smaller uncertainties. This result further validates the SPDS extension for $B_1$ mapping.

The analysis of the data in Figure~\ref{fig:mapasrodillapetra}c, according to Eq.~(\ref{eq:lambdaVFAratio}), provides the estimated $T_1$ values for the different soft and hard tissues in the knee, summarized in the titles of Figure~\ref{fig:mapasrodillapetra}d. The mean $T_1$ values obtained for muscle and fat with VFA-PETRA agree, within experimental error, with those derived from standard methods, confirming the reliability of the proposed approach. However, it was not possible to quantify the $T_1$ times of the smallest short-$T_2$ tissues, such as the meniscus, due to insufficient segmentation accuracy. Conversely, patellar cartilage and the posterior cruciate ligament were included in the analysis despite the more challenging conditions arising from the limited number of voxels contained within their respective masks (28 and 36 voxels, respectively). Consequently, the reported $T_1$ values for these tissues should be interpreted with caution.

To the best of our knowledge, this is the first report of \emph{in-vivo} $T_1$ values for hard tissues at $B_0<0.1$\,T. Nevertheless, we acknowledge the limited precision of these still preliminary results, which are not intended to constitute the central contribution of this work. Importantly, the knee $T_1$ map presented here was not generated from a pair of PETRA sequences specifically optimized for this purpose \cite{Schabel2009Uncertainty}, a topic whose depth and complexity we have chosen to reserve for future studies. Notably, several VFA approaches have already been investigated at $B_0>3$\,T \cite{Liberman2014,Heule2016,Lee2017,Belsley2023}, where the broader dispersion of tissue $T_1$ values and the substantially higher SNR enable the generation of highly precise $T_1$ maps without requiring meticulous optimization or detailed field modeling.

In our case, at $B_0\approx90$\,mT, although the estimated $T_1$ accuracy is high, VFA-PETRA produces $T_1$ maps with insufficient SNR, resulting in imprecise voxel-wise $T_1$ estimations. Furthermore, the intrinsically low SNR of LF-MRI and the need to maintain reasonable scan times impose relatively coarse spatial resolutions, particularly along the slice direction. This introduces segmentation uncertainties that are difficult to resolve, even when employing interpolation techniques such as zero-padding, since hard tissues in the knee can be small structures that would ideally require submillimetric resolution for proper discrimination. Thus for lower fields, while the shortening of $T_1$ values may allow shorter TRs and reduced scan times, $T_1$ values across tissues also become more similar, thereby reducing intrinsic $T_1$ contrast and complicating the optimization of VFA protocols for robust tissue discrimination. In addition, the large voxel size exacerbates partial-volume effects between adjacent tissues, such that the analysis of a given structure inevitably includes contributions from surrounding tissues. On the contrary, lower fields are easier to homogeneize, both in relative (ppm) and absolute terms (Hz), leading to smoother $B_0(\vec{r})$. 

This limitation becomes especially critical when attempting a comprehensive analysis of hard tissues, as the small size of ligaments, tendons, and cartilage substantially restricts the number of voxels available for meaningful statistical analysis, unlike muscle and fat, which occupy much larger regions. Moreover, hard tissues contain significantly lower $^{1}\mathrm{H}$ density than muscle or fat, further reducing SNR and thereby increasing the uncertainty of the estimated $T_1$ values.

Finally, given the low field strength, short RF pulses, and low RF duty cycle of the PETRA acquisition, magnetization transfer effects are expected to be negligible \cite{Su2024}, and the reported $T_1$ values represent apparent relaxation times without the need for explicit two-pool modeling.

\section{Conclusion}
\label{sec:conclusions}

In this work, we have demonstrated that it is possible to broaden the range of pulse sequences applicable to portable, low-cost MRI systems based on Halbach magnets. Specifically, we have implemented and optimized a ZTE-like PETRA sequence capable of producing images with clinically acceptable resolution and SNR within acquisition times below 15\,min at $B_0 \approx 90$\,mT. This achievement significantly extends the diagnostic reach of such portable systems, enabling the visualization of hard tissues (including ligaments, tendons, cartilage, and bone) that are largely suppressed in conventional spin-echo acquisitions. The feasibility of \emph{in-vivo} PETRA imaging was demonstrated on knees and ankles of healthy volunteers, establishing, to our knowledge, the first proof of ZTE imaging in a Halbach-based MRI system. Clinically, this holds promise for musculoskeletal assessment, therapy monitoring, and longitudinal follow-up in both radiological and point-of-care environments, offering a non-ionizing alternative to X-rays for bone and ligament evaluation.

To achieve artifact-mitigated PETRA reconstructions under the strong field inhomogeneities characteristic of Halbach systems, a set of dedicated calibration and correction procedures was developed. These include the use of pre/counter-emphasis pulses to minimize RF ring-down energy and reduce Tx/Rx switching time, thereby improving the temporal efficiency and SNR of the sequence without extending scan duration or computational demand. Furthermore, we introduced an extension of the SPDS protocol that enables simultaneous mapping of $B_0(\vec{r})$ and $B_1(\vec{r})$ in under 5\,min. This calibration, which can be performed entirely with a phantom outside the patient workflow, was found essential for accurate model-based reconstruction, as $B_0$ PK proved critical for mitigating geometric distortions. In contrast, $B_1$ PK, while not mandatory for qualitative imaging, was shown to be crucial for accurate quantitative mapping.

Using these calibrations, we assessed the VFA-PETRA approach for quantitative $T_1$ mapping in both phantoms and \emph{in-vivo} in human knees. This method enables, for the first time, the estimation of $T_1$ in hard tissues with $T_2 < 1$\,ms at $B_0 < 0.1$\,T, extending quantitative MRI capabilities into regimes previously inaccessible to portable scanners. Although the technique apparently provides accurate results at least for soft tissues, its precision remains limited by the intrinsically low SNR of low-field MRI, the coarse spatial resolution required for clinically acceptable scan times, and the small physical dimensions and low proton density of hard tissues. These constraints amplify partial-volume effects and reduce segmentation reliability, particularly for ligaments and cartilage. Similar challenges have been reported in other low-field VFA implementations \cite{Shen2025}, underscoring the need for further optimization of both sequence parameters and hardware sensitivity. Moreover, the limited number of volunteers included in this study warrants caution regarding the reliability and reproducibility of the method, particularly when considering scanners with characteristics different from those of our system. In this regard, we believe that two main aspects must be addressed in future work: first, a careful optimization of the spatial resolution, flip angles, and repetition times to generate more accurate $T_1$ maps within clinically acceptable scan times; and second, the implementation of a systematic, radiologist-supervised clinical study to evaluate the predictive reliability of the method in both injured and healthy knees across subjects with different ages, sexes, and physical conditions.

Despite these limitations, the present work establishes a solid foundation for quantitative and structural MRI at low fields. The demonstrated combination of PETRA imaging, pre/counter-emphasis, and SPDS-based field mapping constitutes a versatile toolbox for robust imaging in portable Halbach systems. Future work will focus on improving SNR through optimized RF coils and reconstruction algorithms, refining VFA-PETRA for higher precision in $T_1$ estimation, and exploring strategies to measure $T_2$ and $T_2^*$ in hard tissues. The latter is currently unfeasible with PETRA but potentially achievable through modified ZTE or multi-echo variants. Together, these advances could further extend the clinical reach of portable MRI, providing safe, accessible, and comprehensive imaging for musculoskeletal applications in both clinical and remote settings.

\section*{Data availability}
All datasets and reconstruction and postprocessing methods generated or used during the present study are available from the corresponding author on reasonable request.

\section*{Ethics statement}
All participants in this work were adults and informed consents to participate and for publication were obtained from the volunteers prior to study commencement. The authors attest that informed consent for publication of identifying information/images in an online open-access publication was obtained from all subjects and/or their legal guardian(s). Ethical approval was obtained from the Ethics Committee of the Spanish National Research Council (research agreement number 276/2025).

\section*{Competing interests policy}
JMA, FG and JA are co-founders of PhysioMRI Tech S.L. TGN consults for PhysioMRI Tech S.L. The remaining authors declare no competing interests.

\section*{Research funding}

This work was funded by: the European Innovation Council (NextMRI 101136407), Ministerio de Ciencia e Innovación (PID2022-142719OB-C22), and the Agència Valenciana de la Innovació under grant INNVA1/2023/30.

\section*{Disclosures}

The tissue identification presented in Figures~\ref{fig:kneeisotropica} and~\ref{fig:rarevspetra} was performed by the authors without input from certified radiological specialists. Consequently, certain structures in the images may have been incorrectly associated with anatomical tissues of the human atlas. It is not our intention to mislead or confuse expert readers, nor to attribute undue interpretative significance to these identifications. For anatomical comparison and tissue labeling, reference was made to the online knee atlas available at  
\href{https://freitasrad.net/pages/atlas/Knee/knee.php}{https://freitasrad.net/pages/atlas/Knee/knee.php}.

\section*{Contributions}

See Table~\ref{tab:contributions}.

\begin{table*}[t]
	\centering
	\caption{Author contributions. An “x” indicates participation in the corresponding task.}
	\label{tab:contributions}
	\begin{tabular}{lccccccccccccccccc}
		\toprule
		\textbf{Task} & \textbf{JB} & \textbf{LGCS} & \textbf{LVC} & \textbf{EC} & \textbf{MFG} & \textbf{PB} & \textbf{RB} & \textbf{JC} & \textbf{PGC} & \textbf{AGC} & \textbf{TGN} & \textbf{EP} & \textbf{LP} & \textbf{LS} & \textbf{JMA} & \textbf{FG} & \textbf{JA} \\
		\midrule
		Scanner preparation & x &   &   &   &   & x & x & x & x &   & x & x & x & x & x &   &   \\
		Sequence design     & x &   &   &   &   &   &   &   &   &   &   &   &   &   & x & x &   \\
		Data acquisition    & x & x & x & x & x &   &   &   & x & x &   &   &   &   &   &   &   \\
		Data analysis       & x & x &   &   &   &   &   &   &   &   &   &   &   &   & x & x & x \\
		Project conception  & x &   &   &   &   &   &   &   &   &   &   &   &   &   & x & x & x \\
		Project management  &   &   &   &   &   &   &   &   &   &   &   &   &   &   &   & x & x \\
		Figure prod.        & x &   &   &   &   &   &   &   &   &   &   &   &   &   &   &   &   \\
		Paper writing       & x &   &   &   &   &   &   &   &   &   &   &   &   &   &   &   & x \\
		Paper revision      & x & x & x & x & x & x & x & x & x & x & x & x & x & x & x & x & x \\
		\bottomrule
	\end{tabular}
	\normalsize
\end{table*}

\bibliography{myrefs}

\end{document}